\documentclass[journal]{IEEEtran}
\usepackage{amsmath,amsfonts}
\usepackage{algorithmic}
\usepackage{algorithm}
\usepackage{array}
\usepackage[caption=false,font=normalsize,labelfont=sf,textfont=sf]{subfig}
\usepackage{booktabs}
\usepackage{textcomp}
\usepackage{stfloats}
\usepackage{url}
\usepackage{verbatim}
\usepackage{graphicx}
\usepackage{cite}

\usepackage{multirow}
\usepackage{threeparttable}

\usepackage{xcolor}
\usepackage[table]{xcolor}
\definecolor{lightblue}{RGB}{220,230,255}
\usepackage[T1]{fontenc}
\usepackage{amsmath}
\usepackage{pifont}
\usepackage{tikz}
\usetikzlibrary{positioning,calc,arrows.meta}
\usepackage{graphicx}
\usepackage{tabularx}
\usepackage{enumitem}

\begin{document}

\title{A Survey of Audio Reasoning in Multimodal Foundation Models}

\author{Zhihan~Guo*,
        Wenqian~Cui*,
        Guan-Ting~Lin,
        Daxin~Tan,
        Jingyao~Li,
        Qiyong~Zheng,
        Dingdong~Wang,
        Jing~Xiong,
        Han~Shi,
        Jiaya~Jia,~\IEEEmembership{Fellow,~IEEE}
        and~Irwin~King,~\IEEEmembership{Fellow,~IEEE}
\IEEEcompsocitemizethanks{\IEEEcompsocthanksitem Zhihan~Guo, Wenqian~Cui, Jingyao~Li, Qiyong~Zheng and Irwin~King are with Department of Computer Science and Engineering of The Chinese University of Hong Kong (CUHK), e-mail: zhguo22@cse.cuhk.edu.hk
\IEEEcompsocthanksitem Guan-Ting~Lin is with the Graduate Institute of Communication Engineering of National Taiwan University
\IEEEcompsocthanksitem Daxin~Tan is with the Department of Electronic Engineering of The Chinese University of Hong Kong (CUHK)
\IEEEcompsocthanksitem  Dingdong~Wang is with the Department of Systems Engineering and Engineering Management of The Chinese University of Hong Kong (CUHK)
\IEEEcompsocthanksitem Jing~Xiong is with the Department of Electrical and Computer Engineering of The University of Hong Kong (HKU)
\IEEEcompsocthanksitem Han~Shi and Jiaya~Jia is with the Department of Computer Science and Engineering of The Hong Kong University of Science and Technology (HKUST)
\IEEEcompsocthanksitem Zhihan~Guo and Wenqian~Cui contributed equally to this work.}}%



\maketitle

\begin{abstract}
Reasoning has become a defining capability of modern foundation models, yet its development in the audio modality remains limited. Audio poses challenges that are distinct from those of text and vision. It is continuous, temporally dense, and contains linguistic, paralinguistic, and environmental information at multiple time scales. As a result, audio reasoning models must align acoustic signals with the discrete semantic space of large language models, while still preserving fine-grained information needed for reliable inference. Progress is also limited by three major obstacles: the scarcity of genuinely audio-grounded reasoning data, shortcut learning and modality hallucination, and the tension between reasoning depth and real-time latency in spoken interaction. In this paper, we present the first dedicated survey of audio reasoning. We provide a unified formulation that distinguishes direct predictive modeling from reasoning-augmented generation, review the architectural and training foundations of audio reasoning models, and systematically organize recent advances in Audio-to-Text, Audio-to-Speech, Audio-Visual Reasoning and Agentic Audio Reasoning. We further examine emerging paradigms such as Chain-of-Thought prompting, supervised fine-tuning, reinforcement learning, and latency-aware spoken interaction, and discuss evaluation practices, open challenges, and future directions. Our goal is to offer a coherent roadmap for developing robust, efficient, and natively grounded audio reasoning systems.
\end{abstract}


\begin{figure*}
    \centering
    \includegraphics[width=1.0\linewidth]{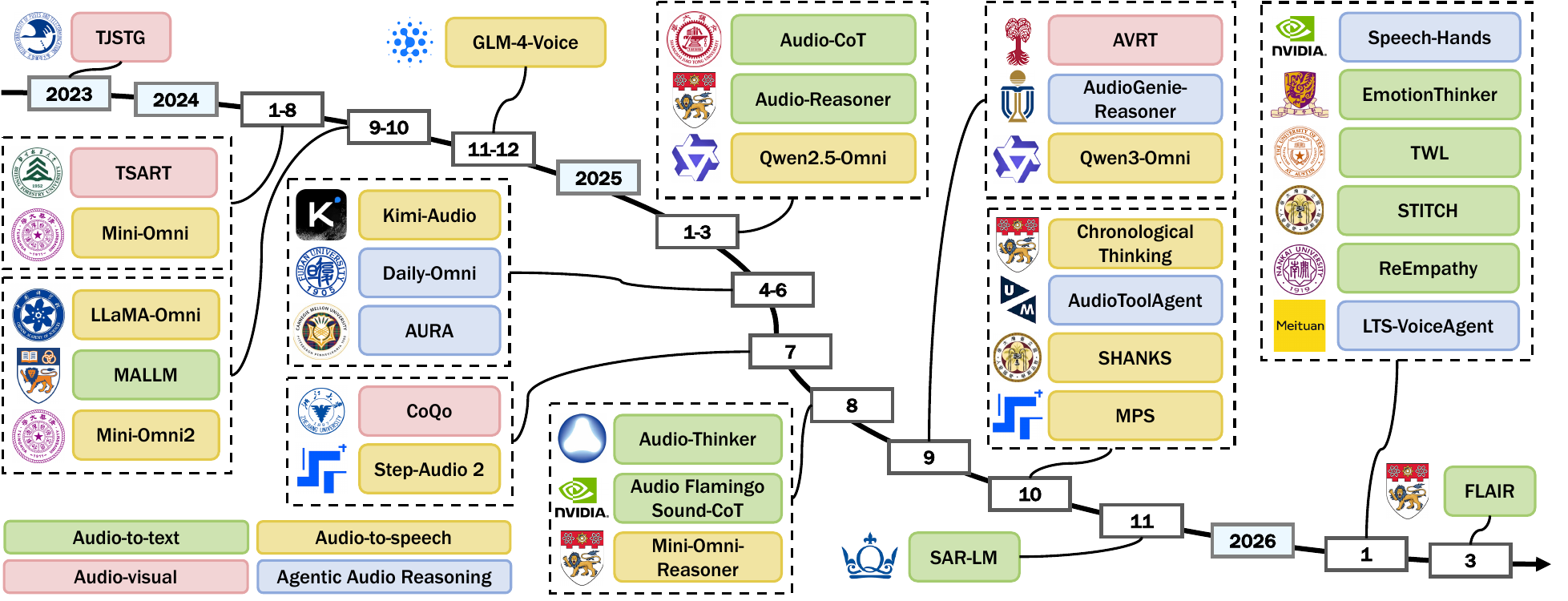}
    \caption{\textbf{Timeline of representative audio reasoning models.} Models are organized chronologically and grouped by major paradigms, including Audio-to-Text, Audio-to-Speech, Audio-Visual, and agentic audio reasoning. 
    }
    \label{fig:audio_reasoning_timeline}
\end{figure*}
\section{Introduction}



Chain-of-Thought (CoT) reasoning has fundamentally reshaped the development of modern foundation models, yielding substantial gains in complex tasks such as mathematical reasoning, coding, and logical problem solving~\cite{wei2022chain, li2025system, guo2025deepseek}. For instance, DeepSeek-R1 achieves superior performance on verifiable tasks such as mathematics and coding competitions, reaching unprecedented accuracy on rigorous benchmarks like AIME and MATH-500~\cite{dekoninck2026matharena, lightman2024let}. Recent reasoning-oriented Large Language Models (LLMs) demonstrate that allocating additional test-time compute can markedly improve performance by enabling longer trajectories of exploration, verification, self-correction, and strategic refinement before producing a final answer~\cite{snell2025scaling, wu2025inference}. This paradigm has also been extended successfully to Vision-Language Models (VLMs), where explicit intermediate reasoning improves performance on tasks requiring dense perceptual grounding, including spatial reasoning, robotic decision making, and medical visual question answering~\cite{shao2024visual, mitra2024compositional}. In contrast, despite the rapid progress of Audio Foundation Models (AFMs), the question of how to endow these systems with robust reasoning capabilities remains largely unresolved~\cite{arora2025landscape, sakshi2025mmau}.

Advancing audio reasoning is important for two reasons. First, audio is not merely an alternative carrier of textual content. It also encodes rich paralinguistic and environmental signals, such as prosody, speaker state, emotion, overlap, temporal dynamics, and background events, that are often essential for reliable inference but are partially or entirely lost in text transcription~\cite{ao2024sd, guo2025recent}. Second, natural human communication is fundamentally speech-centric~\cite{arnon2025enables, wandelt2024representation}. If future AI systems are expected to operate in realistic, embodied, and interactive settings, they must reason directly over acoustic signals and respond in forms compatible with human spoken interaction, rather than relying exclusively on cascaded text-based pipelines~\cite{fang2024llama, zhang-etal-2025-omniflatten}.

However, incorporating reasoning into AFMs is not merely a matter of transplanting CoT techniques from text or vision~\cite{sprague2025to, gao-etal-2025-benchmarking}. Audio reasoning introduces a set of modality-specific research questions that can be organized according to the form of input evidence, output modality, interaction setting, and reasoning workflow. Guided by these dimensions, this survey organizes the field into four paradigms. In Audio-to-Text reasoning, models are expected to infer textual answers from acoustic signals by exploiting linguistic content, paralinguistic cues, environmental events, and temporal dynamics, making acoustic grounding a central requirement. In Audio-to-Speech reasoning, reasoning is embedded in spoken interaction: systems must understand audio inputs, perform intermediate deliberation, and generate speech responses while balancing reasoning depth against conversational latency. In Audio-Visual reasoning, the reasoning process must integrate synchronized auditory and visual evidence, requiring temporal alignment, cross-modal grounding, and robust disambiguation across modalities. Beyond these single-model paradigms, Agentic Audio Reasoning decomposes complex audio-centered tasks into perception, planning, tool use, memory, reflection, and multi-module collaboration, thereby extending audio reasoning from direct generation to structured problem solving. This taxonomy clarifies the scope of audio reasoning and highlights the field’s current fragmentation across formulation, architecture, training, interaction, grounding, and evaluation.

These gaps also expose a limitation of the existing survey literature. Recent reviews have systematically examined audio large language models, spoken language models, multimodal alignment, instruction tuning, real-time speech interaction, and evaluation pipelines~\cite{cui2025recent, arora2025landscape}. In parallel, recent surveys on multimodal Chain-of-Thought discuss reasoning across modalities more broadly~\cite{latif2023sparks}. Nevertheless, in most of these works, reasoning in the audio modality is treated as a secondary capability~\cite{su2025audio, peng2025survey, cui2025recent, arora2025landscape, wang2025towards, yang2025towards, wang2025multimodal}, rather than as the central focus. As a result, the field still lacks a dedicated synthesis that centers audio reasoning itself: how it should be formulated, what architectural and training foundations it requires, why existing reasoning paradigms behave differently in audio than in text, how real-time spoken reasoning changes the design space, and how genuine acoustic grounding should be evaluated. To the best of our knowledge, this work is the first survey specifically devoted to audio reasoning.

To address this need, we present a structured survey of audio reasoning that connects problem formulation, model foundations, reasoning paradigms, grounding, and evaluation. Specifically, our contributions are as follows:

\begin{itemize}
    \item \textbf{A unified formulation of audio reasoning.}
    We define common notations and formalize audio reasoning across text-output, speech-output, audio-visual, and agentic settings under a unified framework.

    \item \textbf{A review of model foundations.}
    We summarize the foundations of audio reasoning models, including audio encoders, modality projectors, LLM backbones, speech tokenization, alignment, instruction tuning, preference optimization, and parameter-efficient adaptation.

    \item \textbf{A comprehensive taxonomy of audio reasoning paradigms.}
    We organize studies into Audio-to-Text, Audio-to-Speech, Audio-Visual, and Agentic Audio Reasoning, covering CoT prompting, supervised fine-tuning, reinforcement learning, latency-aware spoken interaction, cross-modal grounding, and tool-augmented reasoning.

    \item \textbf{A discussion of evaluation and future directions.}
    We summarize evaluation protocols, identify their limitations, and outline open challenges for building robust, efficient, and natively grounded audio reasoning systems.
\end{itemize}

The remainder of this paper is organized as follows. 
Section~\ref{sec: Notations and Problem Formulations} introduces the notations and formal problem formulations for audio reasoning. 
Section~\ref{sec:Foundations for Audio Reasoning Models} reviews the architectural and training foundations of audio reasoning models. 
Section~\ref{sec: Audio-to-Text Reasoning} surveys Audio-to-Text reasoning methods, including CoT prompting, supervised fine-tuning, reinforcement learning, and reasoning-oriented dataset construction. 
Section~\ref{sec: Audio-to-Speech Reasoning} discusses sequential and real-time Audio-to-Speech reasoning, with emphasis on latency-aware interaction paradigms. 
Section~\ref{sec: Audio-Visual Reasoning} examines Audio-Visual reasoning, cross-modal grounding, and audio-visual benchmarks. 
Section~\ref{sec:agentic_audio_reasoning} reviews Agentic Audio Reasoning, including predefined workflow agents, dynamic tool-calling agents, and design patterns. 
Section~\ref{sec: Evaluations} summarizes evaluation settings, benchmarks, and metrics for audio reasoning. 
Finally, Section~\ref{sec: Challenges and Future Directions} discusses open challenges and future research directions, including data reliability, modality hallucination, latency constraints, long-context audio reasoning, and reasoning-oriented pre-training. 

\begin{figure}
    \centering
    \includegraphics[width=1.0\linewidth]{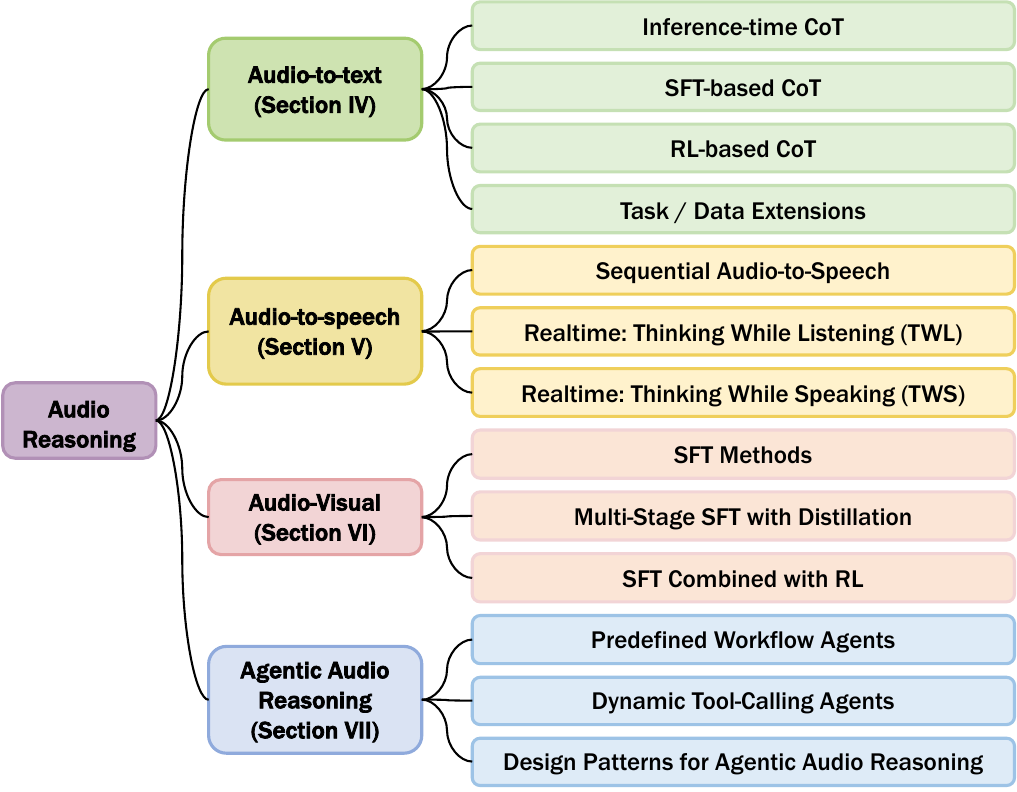}
    \caption{\textbf{A compact taxonomy of audio reasoning.} We organize the literature into four paradigms: Audio-to-Text reasoning, Audio-to-Speech reasoning, Audio-Visual reasoning, and Agentic Audio Reasoning. 
    Representative methods and design patterns are discussed in the corresponding sections.}
    \label{fig:audio_reasoning_taxonomy}
\end{figure}

\section{Notations and Problem Formulations}
\label{sec: Notations and Problem Formulations}

In this section, we establish a unified mathematical framework for the audio reasoning settings discussed throughout this survey. Our goal is not to impose a model architecture, but to provide a consistent notation system that covers the four paradigms reviewed later in the paper: Audio-to-Text reasoning, Audio-to-Speech reasoning, Audio-Visual reasoning, and Agentic Audio Reasoning. We begin with the notational conventions for inputs, intermediate representations, reasoning states, and outputs, and then formalize each reasoning setting under a common probabilistic view.
For clarity, Table~\ref{tab:notation} summarizes the main symbols used throughout the paper.

\subsection{General Formulation of Audio Reasoning}

Audio reasoning can be viewed as conditional generation over multimodal evidence with an optional intermediate reasoning process. 
As summarized in Table~\ref{tab:notation}, we use $\mathbf{A}$, $\mathbf{X}$, and $\mathbf{V}$ to denote audio, textual, and visual inputs, respectively, and use $\mathbf{C}$ to denote the unified context derived from the available modalities. 
In its most general form, the model conditions on $\mathbf{C}$ and produces an output sequence $\mathbf{O}$, optionally through an intermediate reasoning trajectory $\mathbf{R}$:
\begin{equation}
P(\mathbf{O} \mid \mathbf{C})
\quad \text{or} \quad
P(\mathbf{R}, \mathbf{O} \mid \mathbf{C}) 
= P(\mathbf{R} \mid \mathbf{C}) P(\mathbf{O} \mid \mathbf{C}, \mathbf{R}).
\end{equation}
Here, $\mathbf{O}$ can be instantiated as a textual response $\mathbf{Y}$, a speech-token response $\mathbf{S}$, or another task-specific output. 
The key distinction between conventional audio understanding and reasoning-enhanced generation is whether the model explicitly or implicitly constructs an intermediate reasoning trajectory that mediates the mapping from perceptual input to output.

\subsection{Scope of the Formulation}

The general formulation above provides a common abstraction for the specific settings discussed in this survey. 
In Audio-to-Text reasoning, the context is constructed from audio input $\mathbf{A}$ and textual instruction $\mathbf{X}$, and the output $\mathbf{O}$ is instantiated as a textual response $\mathbf{Y}$. 
In Audio-to-Speech reasoning, the output becomes a speech-token sequence $\mathbf{S}$; in real-time settings, streaming audio chunks $\mathbf{A}^{\mathrm{str}}$, stepwise reasoning states $\mathbf{R}_t$, and speech chunks $\mathbf{S}_t$ further characterize the causal interaction among listening, thinking, and speaking. 
In Audio-Visual reasoning, visual input $\mathbf{V}$ is incorporated into the context $\mathbf{C}$, enabling reasoning over jointly grounded acoustic and visual evidence. 
In Agentic Audio Reasoning, the same abstraction defines the perceptual input, intermediate reasoning trajectory, and task output, while agent-specific mechanisms such as planning, tool use, memory, reflection, and multi-module collaboration are discussed in Section~\ref{sec:agentic_audio_reasoning}.

\begin{table}[t]
\centering
\caption{Core notations used in this survey.}
\label{tab:notation}
\footnotesize
\setlength{\tabcolsep}{20pt}
\begin{tabular}{ll}
\toprule
Symbol & Meaning \\
\midrule
$\mathbf{A}=\{\mathbf{a}_t\}_{t=1}^{T_a}$ & Audio input or feature sequence \\
$\mathbf{X}=\{x_n\}_{n=1}^{N}$ & Text instruction or prompt \\
$\mathbf{V}=\{\mathbf{v}_p\}_{p=1}^{P}$ & Visual input sequence \\
$\mathbf{C}$ & Unified multimodal context \\
$\mathbf{R}=\{r_k\}_{k=1}^{K}$ & Reasoning trajectory \\
$\mathbf{Y}=\{y_m\}_{m=1}^{M}$ & Text output \\
$\mathbf{S}=\{s_\ell\}_{\ell=1}^{L_s}$ & Speech-token output \\
$\mathbf{O}$ & Generic output, e.g., $\mathbf{Y}$ or $\mathbf{S}$ \\
$\mathbf{A}^{\mathrm{str}}=\{\mathbf{A}^{(c)}\}_{c=1}^{C_s}$ & Streaming audio chunks \\
$\mathbf{S}^{\mathrm{str}}=\{\mathbf{S}_t\}_{t\geq 1}$ & Streaming speech chunks \\
$\mathbf{R}_t$ & Reasoning state at step $t$ \\
$\mathbf{S}_t$ & Speech chunk at step $t$ \\
\bottomrule
\end{tabular}
\end{table}

\section{Foundations for Audio Reasoning Models}
\label{sec:Foundations for Audio Reasoning Models}

Audio reasoning fundamentally requires systems capable of deciphering the complex, continuous nature of sound and translating these acoustic signals into logical comprehension or expressive generation. The foundation of these advanced cognitive abilities is predicated upon two critical dimensions---This section discusses the \textbf{1) underlying model architectures} designed to effectively bridge audio signals with language modeling, alongside the \textbf{2) training technologies} necessary to cultivate the reasoning abilities.

\subsection{Underlying Model Architectures}

\subsubsection{Large Audio Language Models (LALMs)}
LALMs aim to enable the audio comprehension abilities to the LLMs. They take audio and optional text instructions as input and generate text as output. Their dominant architecture is an encoder--projector--LLM pipeline, where an audio encoder first extracts continuous representations from the input audio, a modality projector maps these representations into the embedding space of a text LLM, and the LLM then performs autoregressive text generation conditioned on the projected audio features and optional textual instructions. In LALMs, audio serves as an external input modality aligned with the textual reasoning space of the LLMs. This design enables efficient reuse of pretrained audio encoders \cite{radford2023robust, hsu2021hubert, chen2023beats, gong2021ast, whisper} and text LLM backbones \cite{touvron2023llama, llama31, qwen2, chiang2023vicuna}, making LALMs modular and extensible.

\subsubsection{Spoken Language Models (SLMs)}
SLMs are foundation models built for spoken dialogue interactions. Similar to LALMs, SLMs can take audio and optional text as inputs, but their outputs are speech representations or waveforms rather than text. Architecturally, they can follow either the encoder--projector--LLM--decoder design that resembles LALMs but adds a speech generation module \cite{zeng2024glm4voice, fang2024llamaomni} after the LLM, or a more unified token-based design \cite{zhang2023speechgpt, defossez2024moshi}, in which continuous speech is first converted into discrete speech tokens and both input and output speech are modeled autoregressively in a shared token space together with text. The latter formulation is more end-to-end, but it also introduces challenges such as long token sequences, multi-codebook token structures, and waveform reconstruction from generated tokens. In short, the main difference is that LALMs align audio to text for text generation, whereas SLMs must additionally support speech generation, requiring an explicit mechanism for producing and decoding speech representations.

\subsection{Training Technologies}
While architectural designs provide the structural foundation for multimodal perception, the emergence of complex reasoning capabilities relies fundamentally on specific training paradigms. In this part, we explore the underlying mechanisms of the core training paradigms that empower audio models with reasoning capabilities.

\subsubsection{Cross-Modal Alignment Pre-training}
Before an LLM can reason over audio, it must first learn to associate acoustic patterns with linguistic meaning \cite{blsp}. The goal of cross-modal alignment pre-training is therefore to bridge the semantic gap between the audio modality and the LLM's native language space. During this stage, the LLM backbone is often frozen, or only lightly updated, to preserve its linguistic and reasoning capabilities, while the bridging components, such as the modality projector in LALMs or the embedding layers for discrete audio tokens in SLMs, are optimized \cite{tang2024salmonn, zhang2023speechgpt}.
This stage typically relies on large-scale audio-text data, often instantiated through tasks such as ASR and audio captioning. 
The audio context may refer either to projected continuous features or to discrete audio tokens, and the model is usually trained with the standard autoregressive cross-entropy loss.

\subsubsection{Post-training}

While cross-modal alignment pre-training establishes a fundamental mapping between acoustic representations and linguistic semantics, this baseline alignment is typically insufficient for robust instruction following, multi-step reasoning, or reliable user interaction. Consequently, the model must undergo post-training---a suite of optimization procedures designed to refine the behavioral priors of the aligned system and elicit the higher-order reasoning capabilities required for complex multimodal tasks.

Post-training typically proceeds in two primary phases: Supervised Fine-Tuning (SFT) and Alignment Optimization. During SFT, the model is trained on curated instruction-response pairs---often enriched with CoT annotations---to learn the structural logic of following complex audio-related commands. While SFT provides a foundation through demonstration, subsequent alignment via Preference Optimization or Reinforcement Learning (RL), such as PPO \cite{ppo} and GRPO \cite{grpo}, further refines the model's behavior. These methods optimize for higher-level objectives such as factual accuracy, reasoning consistency, and faithfulness to the acoustic evidence, ensuring the model's outputs align with human intent and logical rigor.

\section{Audio-to-Text Reasoning}
\label{sec: Audio-to-Text Reasoning}

Audio-to-text reasoning refers to the process of perceiving and analyzing audio signals through step-by-step logical reasoning to draw inferences, with the final output presented in textual form. To ensure accurate reasoning, this approach fully leverages the multiple types of information embedded in audio signals, including linguistic semantic information (e.g., speech content), paralinguistic cues (e.g., intonation, stress, and emotion), and environmental acoustic information (e.g., background sounds and acoustic events).

To formally define Audio-to-Text reasoning, the system receives audio input $\mathbf{A}$ and textual instruction $\mathbf{X}$, and produces a textual response $\mathbf{Y}$. 
In a direct predictive formulation without explicit reasoning, the model optimizes
\begin{equation}
P(\mathbf{Y} \mid \mathbf{A}, \mathbf{X})
=
\prod_{m=1}^{M} P(y_m \mid \mathbf{A}, \mathbf{X}, \mathbf{Y}_{<m}).
\label{eq:autoregressive}
\end{equation}

Reasoning-augmented models instead introduce an intermediate reasoning trajectory $\mathbf{R}$, yielding the joint formulation
\begin{equation}
P(\mathbf{R}, \mathbf{Y} \mid \mathbf{A}, \mathbf{X})
=
P(\mathbf{R} \mid \mathbf{A}, \mathbf{X})\, 
P(\mathbf{Y} \mid \mathbf{A}, \mathbf{X}, \mathbf{R}).
\end{equation}
Applying the chain rule gives the stepwise autoregressive form
\begin{equation}
\begin{aligned}
P(\mathbf{R}, \mathbf{Y} \mid \mathbf{A}, \mathbf{X})
&=
\prod_{k=1}^{K} P(r_k \mid \mathbf{A}, \mathbf{X}, \mathbf{R}_{<k}) \\
&\quad \cdot
\prod_{m=1}^{M} P(y_m \mid \mathbf{A}, \mathbf{X}, \mathbf{R}, \mathbf{Y}_{<m}).
\end{aligned}
\end{equation}


\subsection{Audio-to-Text Reasoning Methods}

The emergence of CoT prompting has fundamentally changed the landscape of Large Language Models (LLMs) by shifting the focus from simple pattern matching to explicit, step-by-step cognitive processing. Although CoT techniques were initially introduced in the text domain, they have now been widely adopted in the field of audio reasoning, serving as a vital technology for its advancement. This section explores how CoT techniques are applied in audio reasoning, categorizing the discussion based on different learning paradigms: inference-time CoT, SFT-based CoT, and RL-based CoT.

\subsubsection{Inference-Time CoT Reasoning Methods}
Inference-time CoT reasoning methods elicit explicit intermediate reasoning steps directly at LLM decoding time, without modifying model parameters. Audio-CoT \cite{audio-cot} is the first study in the field of CoT-based audio reasoning. It presents the first systematic exploration of integrating inference-time CoT reasoning into LALMs to improve their reasoning capabilities across speech, sound, and music modalities. They investigate three inference-time CoT approaches: 
\begin{itemize}
    \item \textbf{Manual-CoT:} This approach utilizes a few-shot learning paradigm featuring hand-crafted exemplars that provide explicit reasoning chains to guide the model toward structured, logical outputs.
    \item \textbf{Zero-Shot-CoT:} This method employs a task-agnostic ``magic prompt'', such as ``Let's think step by step'', to elicit internal reasoning processes without the requirement of pre-defined demonstrations.
    \item \textbf{Desp-CoT:} This technique involves a two-stage process, where the model first generates a descriptive caption or intermediate audio representation, which is then concatenated with the original instruction to facilitate the reasoning process.
\end{itemize}
The authors found that explicit step-by-step reasoning significantly enhances model performance on easy and medium-difficulty tasks, though it can sometimes confuse the models on harder problems. This study systematically proves the feasibility of integrating CoT into the audio reasoning process, laying a solid foundation for future work.
Another inference-time CoT reasoning paper, SAR-LM \cite{sar-lm}, investigates the concept of symbolic audio reasoning, where complex audio signals are converted into simple, human-readable text descriptions before the AI processes them. 

While inference-time CoT methods successfully elicit latent reasoning without the need for additional training, they are fundamentally constrained by the model’s original parameter distribution and the stochastic nature of zero-shot or few-shot prompting. To move beyond these inherent limitations, researchers have increasingly turned to training-based approaches, including SFT- and RL-based methods. This paradigm enables the model to learn explicitly from domain-specific reasoning chains, thereby improving its audio reasoning performance.

\subsubsection{SFT-Based CoT Reasoning Methods}

While inference-time techniques leverage the latent capabilities of pre-trained models, SFT-based CoT methods involve training LALMs on curated datasets that explicitly pair audio inputs with multi-step reasoning chains. This approach moves beyond simple instruction-following by teaching the model the structural logic required to decompose complex acoustic scenes.

The effectiveness of SFT-based CoT reasoning is fundamentally driven by data construction, relying on the thoroughness and precision of the annotated intermediate steps (e.g., planning, captioning, and summarizing). Because the success of these methods depends so heavily on the quality and structure of the underlying data, readers can refer to the Section \ref{sec:audio2text_dataset} for detailed discussions on specific reasoning datasets, such as the datasets in Audio-Reasoner \cite{audio-reasoner}, Audio Flamingo Sound-CoT \cite{AFSoundCoT}, Audio-Cogito \cite{audio-cogito}, and their generation pipelines. 
Ultimately, SFT-based CoT methods are largely driven by data construction---their performance depends heavily on the quality and structure of annotated reasoning chains. Yet even with well-designed data, true reasoning ability relies on whether the model can internalize and generalize the logic rather than merely imitate it. This constraint has prompted a shift toward RL-based approaches, where reasoning can be more directly optimized through tailored training objectives.

\subsubsection{RL-Based CoT Reasoning Methods}
RL-based CoT reasoning methods aim to further enhance audio reasoning ability by directly optimizing the quality, faithfulness, and usefulness of intermediate reasoning steps through the RL process. RL techniques, such as Reinforcement Learning with Verifiable Rewards (RLVR), have been extensively studied in the text domain and have demonstrated strong effectiveness in enhancing the complex reasoning abilities of LLMs. However, it is unclear whether these techniques can be successfully transferred to the audio domain, especially given the continuous and multi-modal nature of audio signals. Recent studies have begun adapting RL frameworks to the audio modality, yielding promising results in complex acoustic reasoning. It is worth noting that, similar to TLM training, SFT is typically employed as a preliminary step before RL optimization, but our primary focus in this section is on RL.

R1-AQA \cite{r1-aqa} is among the first to empirically demonstrate the effectiveness of RL training in the audio reasoning task. They show that RL training leads to better model performance compared to SFT training. However, they raise questions about the usefulness of intermediate CoT tokens in this task. Specifically, they find that while CoT tokens can enhance model performance in an inference-only setting, they do not improve performance when trained with GRPO. This indicates that leveraging CoT tokens in RL training may face domain-specific challenges when applied to the audio domain. Another domain-specific challenge is highlighted by Omni-R1 \cite{omni-r1-audiouseless}, which questions the effectiveness of audio signals in audio reasoning tasks. They find that many audio QA problems in MMAU \cite{mmau} can be solved correctly by relying solely on text information. Furthermore, they discover that text-only RL fine-tuning can enhance audio QA accuracy. Together, these findings emphasize that audio reasoning requires unique considerations, from problem design to model training.

Researchers have identified several key techniques to enhance audio reasoning abilities. The first technique involves understanding \textbf{problem difficulty}. Studies show that it is crucial to categorize problems according to their difficulty levels during RL training, and problems with different difficulty levels are often treated differently. Wijngaard et al. \cite{wijngaard2025data} utilize a single model, Phi-4-mini-instruct \cite{abdin2024phi}, to directly assess the difficulty of various problems. They implement the curriculum training by first performing RL training on the easy data, followed by training on all the data. SARI \cite{sari} leverages the backbone model itself to solve the target problems. The difficulty label is determined by the ratio of correctly answered questions. Then, curriculum learning is utilized as the training method. Omni-AutoThink \cite{omni-autothink} employs three models to determine the problem difficulty: a base model, a strong non-reasoning model, and a strong reasoning model. The reasoning tokens of the easy problems are excluded from the gradient update during RL training. Additionally, Omni-CLST \cite{omni-clst} uses the target model in two training phases---before and after SFT training on the target problems---to address the problem difficulty. Similarly, the reasoning tokens for the easy problems are dropped. Sheng et al. \cite{sheng2025think} propose utilizing the entropy of attention across all audio tokens, suggesting that a problem is considered more difficult if it exhibits higher entropy. They design the reward function to favor the output of more reasoning tokens on harder problems.

The second technique focuses on the \textbf{information type} that the model should prioritize during RL training, allowing it to better reason about the important aspects or properties of the audio input. AudSemThinker \cite{audsemthinker} suggests that conducting additional audio information extraction prior to reasoning can improve a model's audio reasoning capabilities. This involves extracting the auditory semantics of the audio input, which includes details about who, what, how, when, and where, and using this information for reasoning. AudioMCQ \cite{audiomcq} categorizes audio reasoning problems into two groups based on whether the model requires audio information to answer the questions correctly. They conduct experiments by replacing the original audio with 30 seconds of silence, proposing that RL training should only be applied to problems that necessitate audio. Step-Audio-R1 \cite{step-audio-r1} also emphasizes that the reasoning process should be anchored in the audio information. They implement iterative RL training, where they select data containing acoustic feature analysis from the model's rollouts, training the model to iteratively ground its reasoning in the audio data.

\begin{table}[t]
\caption{Types of Rewards used by RL-Based Audio-to-Text Reasoning Methods.}
\label{tab:rewards}
\begin{center}
\scalebox{0.87}{
  \begin{tabular}{lccccc} 
    \toprule
    \textbf{Models} & \textbf{Accuracy}  & \textbf{Consistency}  & \textbf{Format} & \textbf{Length} & \textbf{Quality} \\
    \midrule
    Step-Audio 2       & \textcolor{green}{\ding{51}} & \textcolor{red}{\ding{55}} & \textcolor{red}{\ding{55}} & \textcolor{green}{\ding{51}} & \textcolor{red}{\ding{55}} \\
    Audio-Thinker       & \textcolor{green}{\ding{51}} & \textcolor{green}{\ding{51}} & \textcolor{green}{\ding{51}} & \textcolor{red}{\ding{55}} & \textcolor{green}{\ding{51}} \\
    CESAR       & \textcolor{green}{\ding{51}} & \textcolor{green}{\ding{51}} & \textcolor{green}{\ding{51}} & \textcolor{green}{\ding{51}} & \textcolor{red}{\ding{55}} \\
    Omni-AutoThink       & \textcolor{green}{\ding{51}} & \textcolor{red}{\ding{55}} & \textcolor{green}{\ding{51}} & \textcolor{red}{\ding{55}} & \textcolor{red}{\ding{55}} \\
    Covo-Audio       & \textcolor{green}{\ding{51}} & \textcolor{green}{\ding{51}} & \textcolor{green}{\ding{51}} & \textcolor{red}{\ding{55}} & \textcolor{green}{\ding{51}} \\
    R1-AQA       & \textcolor{green}{\ding{51}} & \textcolor{red}{\ding{55}} & \textcolor{green}{\ding{51}} & \textcolor{red}{\ding{55}} & \textcolor{red}{\ding{55}} \\
    Omni-R1       & \textcolor{green}{\ding{51}} & \textcolor{red}{\ding{55}} & \textcolor{red}{\ding{55}} & \textcolor{red}{\ding{55}} & \textcolor{red}{\ding{55}} \\
    Sheng et al.       & \textcolor{green}{\ding{51}} & \textcolor{red}{\ding{55}} & \textcolor{red}{\ding{55}} & \textcolor{green}{\ding{51}} & \textcolor{red}{\ding{55}} \\
    Wijngaard et al.       & \textcolor{green}{\ding{51}} & \textcolor{red}{\ding{55}} & \textcolor{green}{\ding{51}} & \textcolor{red}{\ding{55}} & \textcolor{red}{\ding{55}} \\
    SARI       & \textcolor{green}{\ding{51}} & \textcolor{red}{\ding{55}} & \textcolor{green}{\ding{51}} & \textcolor{red}{\ding{55}} & \textcolor{red}{\ding{55}} \\
    Omni-CLST       & \textcolor{green}{\ding{51}} & \textcolor{red}{\ding{55}} & \textcolor{green}{\ding{51}} & \textcolor{red}{\ding{55}} & \textcolor{red}{\ding{55}} \\
    AudioMCQ       & \textcolor{green}{\ding{51}} & \textcolor{red}{\ding{55}} & \textcolor{red}{\ding{55}} & \textcolor{red}{\ding{55}} & \textcolor{red}{\ding{55}} \\
    AudSemThinker       & \textcolor{green}{\ding{51}} & \textcolor{red}{\ding{55}} & \textcolor{green}{\ding{51}} & \textcolor{green}{\ding{51}} & \textcolor{red}{\ding{55}} \\
    Step-Audio R1       & \textcolor{green}{\ding{51}} & \textcolor{red}{\ding{55}} & \textcolor{green}{\ding{51}} & \textcolor{red}{\ding{55}} & \textcolor{red}{\ding{55}} \\
    Audio-DeepThinker & \textcolor{green}{\ding{51}} & \textcolor{green}{\ding{51}} & \textcolor{green}{\ding{51}} & \textcolor{red}{\ding{55}} & \textcolor{green}{\ding{51}} \\
    \bottomrule
  \end{tabular}
}
\end{center}
\end{table}

\textbf{Reward design} is another crucial component for successful RL training. Existing papers primarily design rewards based on five important aspects.
\begin{enumerate}
    \item \textbf{Reasoning accuracy:} This reward measures the model's ability to reach the correct final answer, often using objective metrics like exact match.
    \item \textbf{Reasoning consistency:} This reward encourages the model to maintain logical alignment between its intermediate CoT steps and the final output, ensuring that the conclusion is a direct result of the preceding reasoning.
    \item \textbf{Reasoning format:} This reward enforces structural constraints—such as the use of specific tags like \texttt{<think>} and \texttt{<answer>}---to ensure the model produces a clear, parsable separation between its internal deliberation and its final response.
    \item \textbf{Reasoning length:} This reward serves as a regularizer to prevent ``reasoning collapse'' or excessive verbosity, either incentivizing longer chains for complex problems to discourage shortcuts or penalizing redundant tokens to maintain efficiency.
    \item \textbf{Reasoning quality:} This reward evaluates the validity of the reasoning process itself, penalizing logically flawed steps even when they are consistent with the final answer or happen to yield the correct result. It encourages step-by-step soundness and ensures the conclusion is derived through genuinely correct reasoning rather than coincidental or shortcut solutions.
\end{enumerate}
Specifically, Step-Audio 2 \cite{step-audio-2} design reward that controls reasoning length, boosts reasoning quality, and boosts audio perception abilities. Audio-Thinker \cite{audio-thinker} and Audio-DeepThinker \cite{audio-deepthinker} ensure reasoning accuracy, consistency, format, and the quality of the reasoning tokens. CESAR \cite{cesar} includes all types of rewards except the reasoning quality. Omni-AutoThink \cite{omni-autothink} combines the reasoning accuracy and length rewards by assigning high rewards to answers both accurate and concise.


While most papers tackle the standard audio-to-text reasoning task, other papers \textbf{extend the current learning paradigm to more complex tasks}. SoundMind \cite{soundmind} explores the task of Audio Logical Reasoning (ALR)---determining whether a logical conclusion is entailed or not-entailed by spoken premises, requiring a model to perform formal deductive reasoning directly over speech audio. MALLM \cite{mallm} extends the standard audio-to-text reasoning to the multi-audio scenario, where the models are required to reason over multiple separate audio clips to solve the problem.


\subsubsection{Is CoT Really Helping?}
Despite the widespread adoption of CoT techniques across inference-time, SFT-based, and RL-based paradigms, a closer examination of the empirical findings reveals a more nuanced picture. While CoT reasoning has proven transformative in the text domain, its benefits in audio reasoning are neither universal nor guaranteed. Several studies raise important questions about when, and whether, intermediate reasoning chains genuinely contribute to performance gains—or whether they introduce new failure modes unique to the audio modality. For instance, Audio-CoT \cite{audio-cot} demonstrates that explicit step-by-step reasoning can paradoxically hurt performance on harder problems, while R1-AQA \cite{r1-aqa} finds that CoT tokens that are beneficial at inference time fail to improve---and may even hinder performance under GRPO training. Most strikingly, Omni-R1 \cite{omni-r1-audiouseless} reveals that a substantial portion of audio QA benchmarks can be solved by a model that never listens to the audio at all. Collectively, these anomalies indicate that discovering how to effectively and universally leverage CoT to drive genuine audio reasoning abilities remains a critical open challenge.

\subsection{Dataset and Pre-processing for Audio-to-Text Reasoning}
\label{sec:audio2text_dataset}

The development of audio-to-text reasoning capabilities critically depends on the construction of high-quality training data. Unlike standard audio understanding tasks that rely on simple annotations such as sound event labels or captions, reasoning-oriented training requires explicit intermediate reasoning chains that teach models to decompose complex problems step by step. This subsection surveys the data construction and preprocessing strategies employed across the audio-to-text reasoning literature, covering data sources and reasoning chain construction.

\subsubsection{Data Sources and QA Pair Construction}

The majority of existing works repurpose established audio corpora rather than collecting new recordings. AVQA is the most widely adopted source, used by R1-AQA~\cite{r1-aqa}, Audio-Thinker~\cite{audio-thinker}, SARI~\cite{sari}, Omni-CLST~\cite{omni-clst}, CESAR~\cite{cesar}, Omni-R1~\cite{omni-r1-audiouseless}, and Sheng et al.~\cite{sheng2025think}, who extract audio tracks from the original audio-visual dataset and adapt question text by replacing visual references (e.g., ``video'') with auditory ones (e.g., ``audio''). Other commonly leveraged datasets include AudioSet, AudioCaps, Clotho, MusicBench, and VGGSound, which provide diverse audio content spanning environmental sounds, music, and speech. AudioMCQ~\cite{audiomcq} aggregates seven source datasets covering speech (47\%), sound (39.1\%), music (8.1\%), and temporal reasoning (5.8\%), constructing 571k multiple-choice QA samples. AudSemThinker~\cite{audsemthinker} constructs an entirely new corpus by mining YouTube closed captions (213k samples) and processing them through an ensemble of nine specialized audio, image, and video analysis models.

Two paradigms dominate QA pair construction. The first directly reuses existing QA annotations with minimal modification, as in R1-AQA~\cite{r1-aqa} and Audio-Thinker~\cite{audio-thinker}. The second employs LLMs to synthesize new questions and answer choices from audio captions: Omni-R1~\cite{omni-r1-audiouseless} uses ChatGPT to generate multiple-choice questions from multi-perspective audio captions, SARI~\cite{sari} uses Qwen2.5-72B, and AudioMCQ~\cite{audiomcq} uses Qwen3-235B. Audio-Reasoner~\cite{audio-reasoner} adopts a progressive difficulty strategy, generating three questions per audio clip at increasing complexity levels---from factual queries to multi-step reasoning problems---using Google Gemini. For multi-audio scenarios, PolyAudio~\cite{polyaudio} constructs 110k QA pairs spanning 11 multi-audio reasoning tasks (e.g., cross-clip comparison, chronological ordering, multi-audio entailment) using LLM-generated task-specific prompts.

\subsubsection{Reasoning Chain Construction}

The construction of intermediate reasoning chains---critical for SFT-based CoT training---follows several distinct strategies.

\textbf{Single-LLM generation} is the most prevalent approach, where a powerful LLM produces reasoning chains given audio captions, questions, and answers. Audio-Reasoner~\cite{audio-reasoner} uses Gemini to generate four-component chains (planning, captioning, reasoning, and summarizing) for its 1.2M-sample CoTA dataset. SARI~\cite{sari} uses Qwen2.5-72B to produce both structured (four-section) and unstructured (free-form) CoT formats, demonstrating that structured reasoning yields more robust generalization. AudioMCQ~\cite{audiomcq} employs Qwen3-235B to generate three-stage reasoning chains (question type analysis, audio content analysis, answer selection) along with simplified unstructured versions. Audio-Cogito \cite{audio-cogito} uses Qwen3-Omni-Thinking to generate free-form reasoning chains.

\textbf{LLM-ALM collaborative generation} addresses the concern that text-only LLMs may hallucinate audio-specific details. Audio Flamingo Sound-CoT~\cite{AFSoundCoT} proposes a collaborative pipeline where a text LLM and an audio language model jointly construct reasoning chains through either parallel sub-question decomposition (BFS-style) or multi-round interactive conversation (DFS-style), ensuring that intermediate reasoning steps are grounded in actual audio content.

\textbf{Iterative self-distillation} enables the model to progressively improve its own reasoning data. Step-Audio-R1~\cite{step-audio-r1} introduces Modality-Grounded Reasoning Distillation (MGRD), where the model generates candidate reasoning chains that are then filtered for acoustic grounding (i.e., references to perceptual features rather than text descriptions), logical coherence, and answer correctness, progressively shifting reasoning from text-surrogate to genuinely audio-grounded.

\textbf{Domain-specific feature-enriched generation} augments LLM-based CoT construction with specialized audio analysis. EmotionThinker~\cite{emotionthinker} constructs EmotionCoT-35K by first extracting prosodic features---pitch, energy, speaking rate, word-level stress, and intonation patterns---via classical signal processing and specialized models, then using GPT-4o to generate reasoning traces grounded in these acoustic cues.

Table~\ref{tab:a2t-datasets} summarizes the training datasets and construction methods across the surveyed audio-to-text reasoning methods.

\begin{table*}[t]
\caption{Summary of training datasets and construction methods for audio-to-text reasoning.}
\label{tab:a2t-datasets}
\begin{center}
\scalebox{0.98}{
  \begin{tabular}{llllc}
    \toprule
    \textbf{Model} & \textbf{Source Data} & \textbf{CoT Construction} & \textbf{Training} & \textbf{Scale} \\
    \midrule
    Audio-CoT & MMAU (eval only) & Prompt-based (no training) & --- & --- \\
    SAR-LM & MMAU, MMAR (eval only) & Symbolic features (no training) & --- & --- \\
    Audio-Reasoner & AudioSet, AudioCaps, + 7 others & Gemini (structured 4-part) & SFT & 1.2M \\
    AF Sound-CoT & AudioSkills, Clotho-AQA, + 10 others & LLM-ALM collaborative (BFS/DFS) & SFT & 1.24M \\
    Audio-Cogito & AudioSet, Clotho, AudioCaps, + 7 others & Self-distillation (Qwen3-OmniThinking) & SFT & 545k \\
    R1-AQA & AVQA & None (direct answer) & GRPO & 38k \\
    Omni-R1 & AVQA, VGGSound & ChatGPT-generated QA & GRPO & 38k--54k \\
    SARI & AudioSet, MusicBench, MELD, AVQA & Qwen2.5-72B (structured + unstructured) & SFT+GRPO & 42k \\
    Audio-Thinker & AVQA & Qwen2.5-72B (structured 4-part) & SFT+GRPO & 40k \\
    AudioMCQ & 7 datasets (Clotho, AudioCaps, etc.) & Qwen3-235B (3-stage + simplified) & SFT+GRPO & 571k \\
    AudSemThinker & YouTube closed captions & Qwen2.5-72B (2/3-phase semantic) & SFT+GRPO & 213k \\
    Omni-CLST & AVQA, AudSemThinker data & Reused (guided dropout) & SFT+GRPO & 249k \\
    Omni-AutoThink & Audio-Reasoner, + multi-modal & Hierarchical difficulty calibration & SFT+GRPO & 571k \\
    Sheng et al. & AVQA & --- & GRPO & 38k \\
    Wijngaard et al. & AVQA, ClothoAQA, + 4 others & Gemini2.5-Pro (cold-start) & SFT+GRPO & 64k \\
    CESAR & AVQA (augmented) & --- & GRPO & 38k+ \\
    Step-Audio 2 & AudioSet, AudioCaps, + in-house & LLM from mixing recipes & SFT+RL & Large \\
    Step-Audio-R1 & ASR, paralinguistic, + in-house & Self-distillation (MGRD) & SFT+RLVR & 5M \\
    EmotionThinker & IEMOCAP, MELD, Expresso, + 2 others & GPT-4o (prosody-enriched) & SFT+GRPO & 35k \\
    SoundMind & LogiQA 2.0-NLI (TTS synthesis) & DeepSeek-R1 & SFT+RL & 6.4k \\
    PolyAudio & AudioSet, LibriSpeech, + 5 others & LLM (multi-audio tasks) & SFT+GRPO & 160k \\
    Audio-DeepThinker & AVQA, AudioSet, MagnaTagATune, + others & DeepSeek V3.1 & GRPO & 69k \\
    \bottomrule
  \end{tabular}
}
\end{center}
\end{table*}

\section{Audio-to-Speech Reasoning}
\label{sec: Audio-to-Speech Reasoning}

Integrating reasoning capabilities into end-to-end SLMs fundamentally changes how latency, 
context, and logical depth are jointly managed, in contrast 
to traditional cascaded ASR--LLM--TTS pipelines. In this 
survey, we adopt the terminology ``Audio-to-Speech'' to 
precisely delineate the scope of ``Audio-in, Speech-out'' 
interaction. We use ``Audio-in'' rather than ``Speech-in'' 
because modern acoustic encoders can process not only spoken 
utterances but also environmental sounds and paralinguistic 
acoustic cues, making the input modality broader than speech 
alone. Regarding the output modality, we strictly focus on 
``Speech-out'' systems designed for spoken conversational 
interaction\footnote{We exclude systems targeting general audio synthesis or music generation, whose objectives (acoustic fidelity, musicality) differ fundamentally from reasoning-driven conversational interaction.}.

\subsection{Sequential Audio-to-Speech Reasoning}
\label{sec:sequential}

A broad range of prominent end-to-end Audio-to-Speech models, 
such as Qwen2.5-Omni~\cite{qwen25omni}, 
Kimi-Audio~\cite{ding2025kimi}, 
GLM-4-Voice~\cite{zeng2024glm4voice}, 
Llama-omni~\cite{fang2024llama}, 
Mini-Omni~\cite{xie2024mini}, 
Mini-Omni 2~\cite{miniomni2}, and 
SALM-Omni~\cite{salm-omni}, exhibit reasoning capabilities inherited from their text-based LLM backbones. These models can apply reasoning to audio inputs, but their performance is often limited by the modality gap \cite{textproslm} between text and audio and by the lack of audio-specific reasoning supervision.

To optimize reasoning for the audio modality, recent frameworks including Qwen3-Omni~\cite{xu2025qwen3}, Qwen3.5-Omni~\cite{qwen35_omni},
Step-Audio 2~\cite{step-audio-2}, and 
Mimo-Audio~\cite{mimoaudio}, have explicitly incorporated 
audio-related reasoning data into their training pipelines. 
This approach typically involves introducing audio-to-text 
``thinking'' data to enhance the model's comprehension of 
complex acoustic signals, or utilizing reasoning steps to 
generate more contextually appropriate and natural speech 
responses. Particularly, Qwen3.5-Omni~\cite{qwen35_omni} addresses the text-audio performance gap through on-policy distillation, which distills high-quality reasoning from text-conditioned outputs into audio-conditioned responses. OpenS2S~\cite{opens2s} optionally incorporates a static 
thinking stage prior to speech synthesis, where the model 
reasons over the user's emotional cues before producing an 
empathetic spoken response.

We classify this paradigm as \textit{sequential} reasoning 
because the architecture necessitates a discrete, step-by-step 
pipeline. In all the aforementioned systems, the internal 
text-based reasoning or ``thinking'' process must fully 
conclude before the speech synthesis module can initiate 
output generation. While this sequential dependency preserves 
the logical integrity of the generated response, it introduces 
significant latency that constitutes a major bottleneck in 
human-computer dialogue, motivating the low-latency, real-time 
audio-to-speech reasoning frameworks explored in the subsequent 
section.

Formally, a standard non-streaming, sequential formulation models the generation process as a joint probability that can be factorized to highlight the sequential dependency:
\begin{equation}
P(\mathbf{R}, \mathbf{S} \mid \mathbf{A}, \mathbf{X}) 
= 
P(\mathbf{R} \mid \mathbf{A}, \mathbf{X}) 
\cdot 
P(\mathbf{S} \mid \mathbf{R}, \mathbf{A}, \mathbf{X}).
\label{eq:sequential}
\end{equation}
Here, the system first generates an intermediate reasoning sequence $\mathbf{R}$ conditioned on the fully received audio input $\mathbf{A}$ and any accompanying context $\mathbf{X}$. 
Only after $\mathbf{R}$ is entirely derived can the system produce the final spoken output $\mathbf{S}$. 
This factorization explicitly encapsulates the step-by-step constraint: input perception and internal reasoning must fully conclude before speech synthesis begins.

\subsection{Realtime Audio-to-Speech Reasoning}
\label{sec:realtime}

Unlike the non-streaming formulations in 
Section~\ref{sec:sequential}, real-time Audio-to-Speech 
reasoning requires input perception and output generation to 
proceed incrementally and concurrently, imposing strict causal 
and temporal constraints on the reasoning process. Let the 
streaming audio input be defined as:
\begin{equation}
\mathbf{A}^{\mathrm{str}} 
= 
\{\mathbf{A}^{(1)}, \mathbf{A}^{(2)}, \ldots, \mathbf{A}^{(C_s)}\}
\end{equation}
and let the generated spoken output be a sequence of chunks:
\begin{equation}
\mathbf{S}^{\mathrm{str}} 
= 
\{\mathbf{S}_1, \mathbf{S}_2, \dots\}.
\end{equation}
The sequential formulation in Eq.~\eqref{eq:sequential} is 
insufficient for these systems, because the model cannot wait 
for the full input $\mathbf{A}$ or the complete reasoning sequence $\mathbf{R}$. 
Instead, it must dynamically balance reasoning depth against 
strict response latency at each incremental step.

The architectural strategies for incorporating reasoning into 
real-time audio-in speech-out models can be broadly classified 
into two categories based on \emph{when} the reasoning 
computation occurs relative to the user's input and the 
system's output: reasoning that overlaps with the listening 
window (Section~\ref{sec:think_while_listen}), and reasoning 
that overlaps with the speaking window 
(Section~\ref{sec:think_while_speak}).

\subsubsection{Category 1: Thinking While Listening (Reasoning during Input)}
\label{sec:think_while_listen}

Methods in this category share a core premise: the duration of the user's speech provides a natural time window that can be actively repurposed for computation. Instead of treating the listening phase as passive transcription followed by a delayed deliberation step, these systems reason simultaneously while the user is still speaking. Consequently, by the time the user finishes their utterance, the model is already prepared to respond.

Formally, this concurrent process is modeled as:
\begin{equation}
P(\mathbf{R}_t \mid \mathbf{A}^{\mathrm{str}}_{\le t}, \mathbf{X}, \mathbf{R}_{<t}),
\label{eq:think_listen}
\end{equation}
where 
$\mathbf{A}^{\mathrm{str}}_{\leq t}
=
\{\mathbf{A}^{(1)},\ldots,\mathbf{A}^{(t)}\}$. 
This formulation ensures that the reasoning state at any time $t$ relies strictly on the audio prefix observed up to that moment.

Existing approaches in this category generally differ across three key design choices. The first is the \textbf{reasoning format}---how the internal thoughts are represented, such as through free-form text tokens, structured nodes, or latent states. The second is the \textbf{trigger condition}---whether the reasoning runs continuously as the audio streams in, or is completeness-gated by the detection of a full phrase. The third is the \textbf{update granularity}---how frequently the reasoning state is refreshed, which can be processed per-chunk, per-segment, or per-step.

\textbf{SHANKS~\cite{chiang2025shanks}} takes the most direct 
approach: it splits the input speech into fixed-duration 
chunks and, after each chunk, generates a sequence of unspoken 
free-form reasoning tokens that update the model's 
understanding of the partial utterance. Because the reasoning 
state is refreshed every chunk, it can also be used to decide 
barge-in points, letting the model interrupt the user when it 
has enough information to respond.

\textbf{Shih \textit{et al.}~\cite{shih2026can}} shift the 
focus from how to reason to when to reason. They introduce an 
entropy-based metric called Question Completeness ($\zeta$) 
that tracks how much semantic information has accumulated from 
the user's speech so far, and trigger the reasoning chain as 
soon as $\zeta$ indicates the query is sufficiently complete. 
This avoids both waiting for end-of-turn silence and reasoning unnecessarily at every chunk.

\textbf{Chronological Thinking~\cite{wu2025chronological}} 
constrains the reasoning trace to a structured form. As the user speaks, the 
model emits a chain of typed thinking nodes (entity, intent, 
action, knowledge, logic) instead of free-form language, with 
each node attached to the corresponding semantic segment of 
the utterance. The reasoning is scheduled toward the end of the listening window and can be cut off at any moment, so when the user stops speaking the model halts reasoning and responds immediately, without additional latency.

\textbf{FLAIR~\cite{flair}} replaces explicit token-based 
reasoning with latent reasoning. Instead of emitting discrete 
text tokens during listening, the LLM recursively feeds its 
own hidden state from the previous step back as input, so the 
hidden state itself accumulates reasoning information as 
audio arrives. To supervise this implicit process without CoT 
annotations, the authors train against a non-causal expert 
model that has access to the full dialogue, using an ELBO 
objective so that the causal model's latent matches the 
expert's at each step. The latent update reuses the standard forward pass and therefore adds no inference latency. The continuous hidden state can also be smoothly revised when the user's later words contradict earlier ones, at the cost of the reasoning trace that is no longer expressible in natural language.

\subsubsection{Category 2: Thinking While Speaking (Reasoning during Output)}
\label{sec:think_while_speak}

The second category addresses latency during the generation stage. Rather than borrowing time from the user's listening window, these systems take advantage of a hardware-level speed asymmetry: GPUs can generate tokens much faster than the real-time rate at which audio is played back. This speed difference creates a time buffer during playback, allowing the system to pre-compute subsequent reasoning and speech tokens simultaneously. The model jointly predicts the current speech chunk and the next reasoning step:
\begin{equation}
P(\mathbf{S}_t, \mathbf{R}_{t+1} \mid \mathbf{A}, \mathbf{X}, \mathbf{S}_{<t}, \mathbf{R}_{\le t}).
\label{eq:think_speak}
\end{equation}
Here, $\mathbf{S}_t$ denotes the $t$-th outgoing speech chunk and $\mathbf{R}_t$ denotes the reasoning produced in the $t$-th computation window. 
Notably, the generation is conditioned on the fully received audio $\mathbf{A}$ rather than a streaming prefix $\mathbf{A}^{\mathrm{str}}_{\le t}$. 
This reflects the current paradigm of most models in this category, which still operate sequentially during the listening phase---waiting for the user's utterance to conclude before initiating the concurrent thinking-and-speaking output phase. 
The index offset between $\mathbf{S}_t$ and $\mathbf{R}_{t+1}$ captures the asynchronous pipeline: within a single time window, the model emits $\mathbf{S}_t$ while concurrently computing $\mathbf{R}_{t+1}$ for use in subsequent steps.

Current approaches within this category generally differ in two main aspects. The first is \textbf{structural organization}---how reasoning and speech are integrated. Models may alternate reasoning and speech blocks at the chunk level, interleave both token types within a single sequence, or utilize two separate modules running in parallel. The second aspect is the \textbf{reasoning objective}---what the system is actually ``thinking'' about. This reasoning typically focuses either on solving an underlying logical problem or self-reflecting on the quality of the response being generated.

\textbf{STITCH~\cite{chiang2026stitch}} directly implements 
this playback-driven idea. The model is trained to alternate 
between an unspoken reasoning chunk and a spoken response 
chunk: while the $N$-th audio chunk is being played to the 
user, the GPU uses the idle time to generate the next 
reasoning chunk, which then conditions the generation of the 
$(N+1)$-th audio chunk. The architecture remains a single 
monolithic model, and inference latency is hidden behind the 
system's own speech.

\textbf{Mini-Omni-Reasoner~\cite{xie2025mini}} pushes 
interleaving from the chunk level to the token level: silent 
reasoning tokens and spoken response tokens are mixed within 
a single sequence, with each response token immediately 
preceded by the reasoning tokens that support it. This tight 
local alignment allows reasoning to be embedded continuously 
within the speech stream rather than separated into discrete 
blocks, enabling continuous speech generation without 
pausing for thought. The design relies on a hierarchical 
Thinker-Talker architecture in which only response tokens 
are passed from the Thinker to the Talker for speech 
synthesis.

\textbf{Mind-Paced Speaking (MPS)~\cite{wu2025mind}} 
decouples reasoning and articulation into two separate 
modules. A Formulation Brain produces high-level logical 
thoughts while an Articulation Brain converts thoughts into 
fluent speech; crucially, the Articulation Brain starts 
speaking from preliminary thoughts before the full reasoning 
chain is finished, and the Formulation Brain acts as a 
pacemaker that keeps feeding refined logic to the speaker. 
Because reasoning and articulation run in parallel on 
separate modules rather than taking turns within a single 
model, the speaker can begin responding as soon as 
preliminary thoughts are available.

\textbf{ReEmpathy~\cite{jia2026reflecting}} uses the same thinking-while-speaking architecture but for reflective rather than logical reasoning. The model interleaves fixed-length 
spoken response chunks with free-form reasoning chunks, but 
unlike STITCH's forward-looking deliberation, each reasoning 
chunk is a self-critique on the empathetic quality of what 
has been said, and the next response chunk is 
adjusted accordingly. This shows that the 
thinking-while-speaking architecture extends beyond logical 
derivation to iterative refinement of affective dimensions.

\subsection{Training Data for Spoken Reasoning}
\label{sec:datasets}

\textbf{Reasoning Domains.} Across the surveyed models, spoken 
reasoning has been applied to two primary domains. The dominant 
focus is \textit{mathematical and logical reasoning}, pursued 
by MPS, STITCH, Mini-Omni-Reasoner, and SHANKS, where models 
are evaluated on TTS-synthesized versions of text benchmarks 
such as GSM8K. These studies consistently demonstrate that 
internal chain-of-thought (CoT) is essential for multi-step 
arithmetic in the spoken modality, as the sequential nature 
of mathematical derivation aligns naturally with discrete 
reasoning-token generation. A second, emerging domain is 
\textit{paralinguistic and empathetic reasoning}, which 
targets the prosodic and affective dimensions of communication 
rather than propositional content. ReEmpathy interleaves reflective reasoning 
chunks with spoken output to iteratively refine empathetic 
quality. This divergence reflects an unresolved trade-off 
between responsiveness and the depth of affective reasoning 
required for genuinely empathetic output. Beyond these two foci, FLAIR extends spoken reasoning to open-domain instruction-following and full-duplex conversational QA, showing that latent reasoning can generalize beyond structured mathematical derivation.

\textbf{Data Scarcity and Adaptation.} A primary challenge in 
training reasoning-capable SLMs is the scarcity of native 
spoken reasoning datasets. Consequently, current research 
predominantly relies on adapting text-based benchmarks 
(e.g., GSM8K, MultiWOZ) into spoken form via text-to-speech 
(TTS) synthesis, coupled with reasoning supervision distilled 
from strong text LLMs. The specific adaptation strategy, 
however, varies considerably across architectures.

\textbf{Modality-Specific Pre-processing.} For models such as 
Chronological Thinking, unstructured CoT traces are parsed 
into structured thinking nodes (entity, intent, action, knowledge, and logic) to 
align with the temporal progression of speech. Mini-Omni-Reasoner 
and STITCH instead adopt a text-audio interleaving strategy, 
where reasoning remains as text tokens but is inserted between 
synthesized speech tokens in the training data; this requires 
careful alignment to ensure that ``thought'' tokens are 
positioned immediately before their corresponding audio 
generation triggers. For affective reasoning, ReEmpathy 
generates ``reflection'' tokens that explicitly critique the 
emotional appropriateness of a response, which are then 
interleaved with the dialogue history to supervise the model's 
internal state during inference.
A notable exception is FLAIR, which eliminates the need for 
explicit reasoning annotations altogether: its latent targets 
are produced on-the-fly by a non-causal expert model during 
training via ELBO optimization, sidestepping the dataset 
construction burden that the methods above share.

Table~\ref{tab:realtime_datasets} summarizes the datasets 
utilized across the surveyed literature. We use the notation 
``(S)'' to indicate that the original text dataset has been 
converted into spoken form via TTS synthesis.

\begin{table*}[htbp]
\caption{Comparison of Datasets and Tasks in Realtime 
Audio-to-Speech Reasoning Models. ``(S)'' indicates 
TTS-synthesized versions of originally text-based datasets.}
\label{tab:realtime_datasets}
\begin{center}
\scalebox{0.85}{
  \begin{tabular}{lll}
    \toprule
    \textbf{Model} & \textbf{Domain} & \textbf{Source Data} \\ 
    \midrule
    \multicolumn{3}{l}{\textit{Category 1: Thinking While Listening}} \\
    \midrule
    Chronological Thinking~\cite{wu2025chronological} & Task-oriented Dialogue, Open-domain QA & GenConv(S), SpokenWOZ(S), LlamaQ(S)  \\ 
    SHANKS~\cite{chiang2025shanks} & Math Reasoning & Tulu3-Persona-Math-Grade(S) \\
    Shih \textit{et al.}~\cite{shih2026can} & Common Sense \& Math Reasoning & COT COLLECTION~\cite{cot_collection}(S) \\
    FLAIR~\cite{flair} & Open-domain QA & Synthetic dialogues (GPT-OSS/Qwen2.5/Llama3.1) \\
    \midrule
    \multicolumn{3}{l}{\textit{Category 2: Thinking While Speaking}} \\
    \midrule
    MPS~\cite{wu2025mind} & Math Reasoning, Complex Conversation & Not disclosed \\
    STITCH~\cite{chiang2026stitch} & Math Reasoning, Knowledge QA & VoiceAssistant400K, Tulu-3-Persona(S), Natural Questions(S), TriviaQA(S) \\
    Mini-Omni-Reasoner~\cite{xie2025mini} & Math Reasoning & SPOKEN-MATH-PROBLEMS-3M \\
    ReEmpathy~\cite{jia2026reflecting} & Empathetic Dialogue \& Reflection & EmotionTalk, OpenS2S subset \\
    \bottomrule
  \end{tabular}
}
\end{center}
\end{table*}

\subsection{Discussion on Reasoning with Latency Constraints}
\label{sec:discussion}

While all surveyed methods aim to reduce the initial response 
latency (Time-To-First-Token, or TTFT), experimental results 
reveal substantially different overall performance profiles depending 
on how reasoning and audio generation are temporally 
synchronized. 

\textbf{Thinking While Listening} can effectively eliminate 
additional wait times after the user finishes speaking. For 
example, Chronological Thinking restricts reasoning to the 
duration of the user's utterance, enabling near-instant 
responses. However, this paradigm is fundamentally limited 
by the duration of the user's utterance. If the query is 
brief but requires deep reasoning (e.g., a short mathematical 
question), the listening window may be insufficient to 
complete the necessary computation. More critically, because 
reasoning begins before the user finishes speaking, the model 
operates on incomplete information. This can lead to premature 
or erroneous reasoning---for instance, committing to an 
incorrect problem interpretation if the user's final words 
substantially alter the underlying condition---resulting in 
wasted computation and degraded response quality.
Latent-reasoning variants such as FLAIR partially mitigate 
these issues by maintaining a continuous hidden state rather 
than committing to discrete textual hypotheses, allowing 
smoother revision as new audio arrives. This flexibility 
comes at the cost of interpretability, as the reasoning 
trace is no longer expressible in natural language.

\textbf{Thinking While Speaking} is not constrained by the 
user's utterance length but is instead limited by the ratio 
of reasoning throughput to audio playback speed. Models like 
STITCH and Mini-Omni-Reasoner hide inference latency by 
ensuring that reasoning for the $(N+1)$-th audio chunk 
completes before playback of the $N$-th chunk finishes. 
However, this approach encounters difficulties with highly 
complex tasks. When reasoning chains exceed the available 
playback buffer, the audio generation may stall, resulting 
in perceptible pauses or disfluencies in the output speech.

Each of these paradigms addresses a different facet of the 
latency challenge, and neither is universally optimal. To 
overcome their respective limitations, future systems could 
benefit from a hybrid architecture that combines both 
strategies. A key research question is how to design a 
\textit{reasoning scheduler} that dynamically selects the 
appropriate strategy based on real-time signals---for example, 
determining whether a query is sufficiently complete to 
trigger early reasoning, or whether the reasoning budget fits 
within the available playback buffer. Further open questions include graceful recovery from premature reasoning and computation allocation across parallel reasoning tracks when both strategies run concurrently. Finally, we note that recent work has explored offloading reasoning to an external text-based back-end running asynchronously alongside the speech model (e.g.,~\cite{kuroki2025kame, chien2026moshirag}). While these systems fall outside the scope of this survey, as the speech model itself does not perform reasoning, they represent a complementary architectural direction that could be combined with the internal reasoning paradigms discussed above.

\begin{figure*}[htbp]
    \centering
    \includegraphics[width=0.95\linewidth]{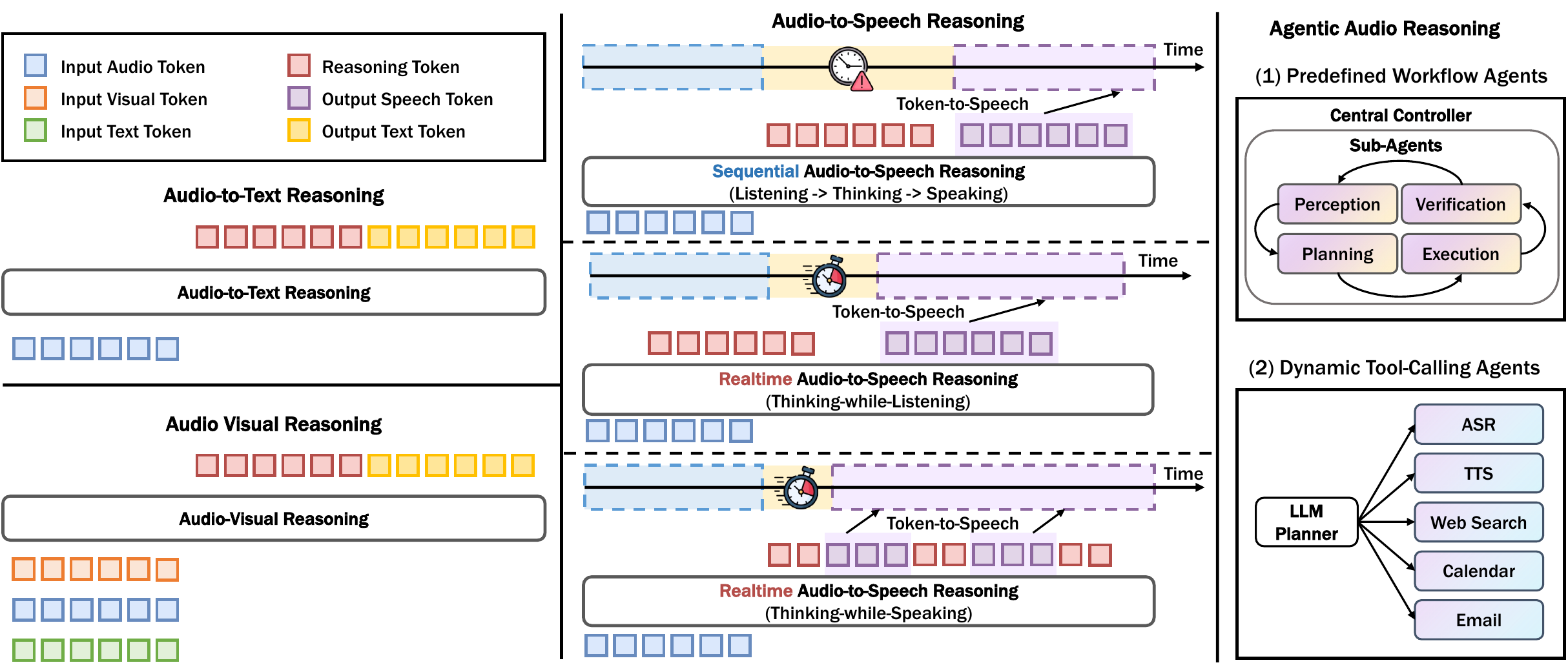}
    \caption{\textbf{Overview of major audio reasoning paradigms.} 
    The figure summarizes four paradigms covered in this survey: Audio-to-Text, Audio-to-Speech, Audio-Visual, and Agentic Audio Reasoning. 
It contrasts text-output reasoning, cross-modal audio-visual grounding, sequential and real-time speech-output reasoning, and agentic workflows based on predefined pipelines or dynamic tool calling.}
    \label{fig:audio_to_speech}
\end{figure*}

\section{Audio-Visual Reasoning}
\label{sec: Audio-Visual Reasoning}

Audio-visual reasoning requires a model to jointly perceive synchronized or loosely aligned acoustic and visual streams and produce a coherent textual response. Unlike unimodal tasks, it demands that evidence from both modalities be integrated, disambiguated, and weighted against a natural-language instruction.

\subsection{Formulation of Audio-Visual Reasoning}

Given audio input $\mathbf{A}$, visual input $\mathbf{V}$, and textual instruction $\mathbf{X}$, a direct model predicts the textual output $\mathbf{Y}$ as
\begin{equation}
P(\mathbf{Y} \mid \mathbf{A}, \mathbf{V}, \mathbf{X})
=
\prod_{m=1}^{M} P(y_m \mid \mathbf{A}, \mathbf{V}, \mathbf{X}, \mathbf{Y}_{<m}).
\label{eq:direct}
\end{equation}
Reasoning-augmented models introduce an intermediate reasoning trajectory $\mathbf{R}$, yielding
\begin{equation}
P(\mathbf{R}, \mathbf{Y} \mid \mathbf{A}, \mathbf{V}, \mathbf{X})
=
P(\mathbf{R} \mid \mathbf{A}, \mathbf{V}, \mathbf{X})\, 
P(\mathbf{Y} \mid \mathbf{A}, \mathbf{V}, \mathbf{X}, \mathbf{R}),
\label{eq:rag}
\end{equation}
which expands autoregressively as
\begin{equation}
\begin{aligned}
P(\mathbf{R}, \mathbf{Y} \mid \mathbf{A}, \mathbf{V}, \mathbf{X})
&=
\prod_{k=1}^{K} P(r_k \mid \mathbf{A}, \mathbf{V}, \mathbf{X}, \mathbf{R}_{<k}) \\
&\quad \cdot
\prod_{m=1}^{M} P(y_m \mid \mathbf{A}, \mathbf{V}, \mathbf{X}, \mathbf{R}, \mathbf{Y}_{<m}).
\end{aligned}
\label{eq:rag_expand}
\end{equation}
Direct models must implicitly resolve audio-visual correspondence within the backbone, leaving no explicit trace of modality weighting. 
Reasoning-augmented models externalize this process as intermediate steps $\mathbf{R}=(r_1,\ldots,r_K)$, improving interpretability and accuracy on multi-hop tasks such as speaker identification or cross-modal event description. 
Current systems represent $\mathbf{R}$ as natural-language thinking tokens~\cite{qwen3omni}, structured sub-questions, or latent intermediate representations, and differ further in whether audio-visual features are fused before (early fusion) or after (late fusion) the LLM decoder.

\subsection{Audio-Visual Reasoning Methods}

We organise existing methods by their post-pretraining optimisation strategy: (i) pure SFT, (ii) multi-stage SFT with distillation, and (iii) SFT followed by reinforcement learning.

\subsubsection{Supervised Fine-Tuning (SFT) Methods}

Most omni-modal models adopt multi-stage SFT pipelines that progressively align modalities before joint instruction tuning. \textbf{VITA}~\cite{fu2024vita} extends a Mixtral-8$\times$7B backbone with modality alignment (InternViT-300M for vision, Whisper-large-v2 for audio) followed by joint multimodal instruction tuning, using state tokens to distinguish noise, audio instructions, and plain text. \textbf{Mini-Omni2}~\cite{miniomni2} adds a SNAC-based parallel text--speech decoding head and a semantic interruption mechanism to a compact 0.5B backbone. \textbf{Megrez-Omni}~\cite{veomni} adopts a three-stage alignment pipeline with a perceiver resampler for vision and joint omni instruction tuning on mixed-modality data. \textbf{Baichuan-Omni}~\cite{baichuan} progressively unfreezes components across four stages and adds a flow-matching speech decoder trained separately from the understanding pipeline. \textbf{Lyra}~\cite{lyra} augments a Qwen2-VL backbone with three components: a Latent Cross-Modality Regularizer (LCMR) that aligns speech and text embeddings via Dynamic Time Warping, modality-specific LoRA adapters to prevent catastrophic forgetting, and a block-wise top-$k$ token extractor to reduce computation on long audio-visual inputs.

\subsubsection{Multi-Stage SFT with Distillation}

\textbf{Qwen2.5-Omni}~\cite{qwen25omni} adopts a Thinker-Talker architecture: the Thinker processes interleaved text, audio, image, and video tokens, while the Talker is a streaming speech decoder refined with DPO~\cite{dpo} for speech quality and multi-speaker instruction tuning. \textbf{Baichuan-Omni-1.5}~\cite{baichuanomni15} constructs cross-modal training data by synthesising text segments into audio via TTS, yielding interleaved image-audio-text prompts that improve audio-visual grounding. \textbf{LongCat-Flash-Omni}~\cite{longcatflashomni} is a large MoE model trained progressively from text-only to full audio-visual alignment, supporting streaming interaction and long-range cross-modal reasoning within a 128K-token context window.

\subsubsection{SFT Combined with Reinforcement Learning (SFT+RL)}

\textbf{Qwen3-Omni}~\cite{qwen3omni} combines SFT, strong-to-weak distillation, and Group Sampling Policy Optimisation (GSPO)~\cite{gspo} with rule-based rewards for verifiable tasks and model-based rewards for open-ended tasks. Its dedicated AuT audio encoder provides richer representations than Whisper-based encoders, and the RL-based thinking mode enables multi-step cross-modal reasoning not achievable by direct-decoding baselines.

\subsection{Audio-Visual Reasoning Benchmarks}
Recent benchmarks for audio-visual reasoning have increasingly shifted beyond visual-dominant evaluation toward scenarios where auditory understanding and temporal alignment are indispensable. \textbf{Daily-Omni}~\cite{zhou2025dailyomni} specifically evaluates whether models can reason over synchronized multimodal streams in everyday settings, emphasizing temporal consistency between acoustic and visual events rather than static scene recognition alone. Similarly, \textbf{WorldSense}~\cite{hong2026worldsense} targets real-world omnimodal understanding, measuring a model’s ability to integrate heterogeneous sensory evidence under realistic conditions. \textbf{AVUT}~\cite{yang2025avut} further highlights audio-centric reasoning by constructing video understanding tasks that remove textual shortcuts, forcing models to rely on sound cues and their correspondence with visual dynamics. Beyond event-level grounding, \textbf{AV-SpeakerBench}~\cite{nguyen2025avspeakerbench} focuses on speaker attribution and human speech understanding, where models must jointly analyze lip movements, facial behaviors, and speaker characteristics in the audio track to determine who is speaking. At a finer semantic level, \textbf{Omni-Cloze}~\cite{ma2025omnicaptioner}, introduced in Omni-Captioner, adopts a cloze-style evaluation protocol to assess detailed multimodal caption understanding, requiring inference over timestamps, entity relations, and subtle auditory details from partially observed descriptions. Collectively, these benchmarks demonstrate that progress in audio-visual reasoning depends not only on visual perception, but increasingly on robust auditory modeling, temporal synchronization, and deep cross-modal integration.

\section{Agentic Audio Reasoning}
\label{sec:agentic_audio_reasoning}

Traditional audio reasoning systems, including audio-to-text, speech-to-speech, and audio-visual reasoning, are commonly built upon monolithic architectures with single-pass inference. While effective for routine tasks, they struggle in complex, open-ended scenarios that require multi-step deduction, ambiguity resolution, error correction, real-time interaction, or cross-modal alignment. To address these limitations, a growing line of research explores \emph{agentic} extensions to audio reasoning, in which autonomous agents perform perception, planning, and execution while flexibly integrating speech, text, and vision modalities. Unlike conventional systems, agentic models perceive inputs, plan strategies, decompose tasks, use tools, reflect on intermediate outputs, and execute multi-stage cross-module reasoning trajectories. In this section, we provide a unified overview of such systems, categorize them by their decision-making workflow, summarize representative works, and discuss their key design patterns.

\subsection{Core Capabilities of Agentic Audio Reasoning}
Agentic systems for audio reasoning typically exhibit five core capabilities:

\begin{itemize}
    \item \textbf{Task decomposition}: Breaking down a complex instruction into smaller, manageable sub-tasks (e.g., audio feature extraction, semantic understanding, logical deduction) to reduce inference complexity.
    \item \textbf{Self-reflection}: Evaluating intermediate outputs to detect and correct errors or resolve ambiguities caused by low-quality signals.
    \item \textbf{Tool use}: Leveraging external modules (e.g., speech recognition, calculators, web search) to overcome a single model's inherent limitations.
    \item \textbf{Adaptability}: Dynamically adjusting the reasoning strategy based on the input audio's characteristics or intermediate findings.
    \item \textbf{Interpretability}: Producing interpretable reasoning traces or explicit chain-of-thought, making the system's decision process transparent and auditable.
\end{itemize}

\subsection{Taxonomy of Agentic Paradigms}

As illustrated in Figure~\ref{fig:audio_to_speech}, we categorize existing agentic systems into two fundamental paradigms based on how the system's core decision flow is defined: \textbf{Predefined Workflow Agents} and \textbf{Dynamic Tool-Calling Agents}.


\subsubsection{Predefined Workflow Agents}
Systems in this paradigm follow a \textbf{fixed, predetermined execution pipeline}. The sequence of stages—such as perception, planning, execution, and verification—is statically designed. A central controller executes these stages in a rigid order without dynamically selecting from a pool of heterogeneous tools at runtime. This closed and predictable structure ensures strong coherence, interpretability, and stability, making it ideal for tasks where reliability is important. However, this fixed structure makes it difficult to extend or adapt to new requirements, and performance is ultimately bounded by the capacity of the core model or the fixed design of its internal sub-agents.

An early example is \textbf{Speech-Hands}~\cite{wan2026speech}, which addresses the problem of an omni-model being misled by noisy inputs. It adopts a three-stage decision pipeline: generating two candidate hypotheses, predicting a special action token (\textit{<internal>}, \textit{<external>}, \textit{<rewrite>}) to choose or rewrite the answer, and producing the final output. All reasoning is confined within a single model, where self-reflection serves as a fixed step instead of a runtime adaptive choice.

A more elaborate multi‑agent design is \textbf{AudioGenie-Reasoner (AGR)}~\cite{rong2025audiogenie}. It runs a training‑free, fixed loop that starts from a coarse transcript and refines it through four specialized agents (\textit{Planning}, \textit{Interaction}, \textit{Augmentation}, \textit{Answering}). The first three agents collaborate to identify gaps and retrieve missing information using tools like ASR or audio QA, and the final agent outputs the answer. The entire sequence is predetermined, yet the iterative refinement enables deep reasoning without parameter updates.

Streaming scenarios challenge the “listen‑then‑act” assumption. \textbf{LTS-VoiceAgent}~\cite{zou2026lts} tackles the “think vs. speak” dilemma with a fixed streaming workflow: a \textit{Dynamic Semantic Trigger} detects meaningful speech boundaries, after which a \textit{Dual-Role Stream Orchestrator} runs a background \textit{Thinker} and a foreground \textit{Speaker} in parallel, allowing speculative generation while listening. The roles are pre‑assigned and do not change at runtime, preserving determinism while reducing interaction latency.

Finally, cross‑modal reasoning appears in \textbf{Daily-Omni Agent}~\cite{DailyOmni}, a training‑free diagnostic baseline. It follows a static pipeline: segment videos and audios, label each modality independently with off‑the‑shelf unimodal models (Qwen2.5-VL for vision, Qwen2-Audio and Whisper-Large-V2 for audio), locate timestamps of critical events, and then generate the answer. All modalities are integrated under a centralized controller, demonstrating that even simple fixed workflows can yield robust audio‑visual reasoning.

\subsubsection{Dynamic Tool-Calling Agents}
In this paradigm, a central planner (typically an LLM) decides \textbf{at runtime} which external modules to call, in what order, and how to combine their outputs. This design emphasizes flexibility, scalability, and modularity, but introduces coordination overhead and potential output inconsistency due to the non-deterministic nature of LLM decisions.

A foundational system is \textbf{AURA}~\cite{maben2025aura}, the first open‑source speech‑native assistant. It implements a ReAct‑style loop where the LLM planner alternates between generating \textit{Thought} and \textit{Action} steps; the \textit{Action} can invoke heterogeneous APIs (calendar, search, email) and the results become \textit{Observation} for the next cycle. This dynamic tool scheduling enables complex, multi‑turn tasks while keeping the assistant architecture fully modular.

A natural extension is \textbf{AudioToolAgent}~\cite{wijngaard2025audiotoolagent}, which coordinates audio‑language models as tools. A central LLM decides which specialist QA models (e.g.,  Qwen2.5 Omni) or ASR modules (e.g., Whisper) to call, how to refine queries, and how to reconcile conflicting outputs. The runtime selection of tools allows it to adapt to diverse audio understanding tasks.

Rather than relying on prompt‑based heuristics, \textbf{AuTAgent}~\cite{tong2026autagent} learns the tool selection policy via reinforcement learning. It combines Group Relative Policy Optimization (GRPO) with a \textit{Baseline‑subtracted Differential Reward Function}, training the agent to invoke external tools only when they yield a net performance gain. The learned policy improves accuracy over both open‑source and closed‑source backbones on audio reasoning benchmarks.

While the above agents operate in a turn‑based or offline fashion, \textbf{SHANKS}~\cite{chiang2025shanks} brings dynamic tool calling to the streaming setting. An SLM generates \textit{unspoken chain‑of‑thought reasoning} while listening to the user. This hidden reasoning drives runtime decisions: whether to interrupt the user (e.g., when a mistake is detected) or to issue tool calls before the user finishes speaking. Despite its streaming nature, each chunk still triggers a fixed CoT generation step, but the tool invocation decision itself is made dynamically.

Another streaming solution, \textbf{Stream RAG}~\cite{arora2025stream}, directly predicts tool queries in parallel with the user’s ongoing speech. Instead of waiting for utterance completion, its central LLM issues tool calls early and incorporates retrieved results into a spoken summary. This preemptive, runtime orchestration reduces tool‑use latency while substantially improving QA accuracy.

Finally, \textbf{VoxMind}~\cite{liang2026voxmind} demonstrates that dynamic tool calling can be embedded in an end‑to‑end, non‑cascaded spoken dialogue model. It introduces a “Think‑before‑Speak” mechanism that internalizes reasoning before response generation, and an asynchronous multi‑agent tool manager that offloads retrieval tasks to a separate agent, decoupling inference latency from the size of the toolset. The tool manager is invoked at runtime based on dialogue context, keeping the model within the dynamic tool‑calling paradigm while achieving state‑of‑the‑art task completion.

\begin{table*}[t]
\centering
\small
\caption{Key trade-offs between predefined workflow and dynamic tool-calling agents.}
\label{tab:tradeoffs}
\begin{tabular}{p{3cm} p{6cm} p{6cm}}
\toprule
\textbf{Dimension} & \textbf{Predefined Workflow} & \textbf{Dynamic Tool-Calling} \\
\midrule
Control flow & Fixed at design time & Decided at runtime by LLM \\
Extensibility & Low (requires redesign) & High (plug-and-play tools) \\
Interpretability & High (predictable traces) & Medium (depends on LLM's reasoning chain) \\
Latency & Low (streaming) to medium (non-streaming) & Medium to high (depends on tool calls) \\
Data/training needs & Low (training optional) & Medium (prompt-based or RL) \\
Suitable scenarios & High-precision, structured tasks & Open-domain, multi-tool collaboration \\
\bottomrule
\end{tabular}
\end{table*} 

\subsection{Design Patterns for Agentic Audio Reasoning}

Beyond the aforementioned two-paradigm taxonomy, we identify representative design patterns that underpin agentic behavior in audio reasoning. In practice, these patterns are non-exclusive and are often integrated within a single system.

\begin{itemize}
\item \textbf{Correction Pattern}: The agent verifies, judges, and corrects its own perception errors in a closed loop. Speech-Hands and Daily-Omni Agent are prime examples, the former using a special token to choose the best hypothesis, the latter using video as ground truth to correct audio predictions. This pattern focuses on improving the \textit{reliability} of the core model.

\item \textbf{Iterative Refinement Pattern}: The agent starts with a low-resolution representation and iteratively enriches it. AudioGenie-Reasoner embodies this, starting with a coarse transcription and refining it through multiple specialized agents, enabling deep reasoning without extra training processes.

\item \textbf{Reactive Tool-Use Pattern}: The agent operates in a \textit{Thought} → \textit{Action} → \textit{Observation} loop, where the action is a tool call. AURA and AudioToolAgent are representative examples, dynamically invoking APIs or specialist models to solve tasks.

\item \textbf{Proactive Streaming Pattern}: The agent takes action \textit{before} it has complete information to reduce latency. SHANKS (CoT generation), LTS-VoiceAgent (speculative speaking), and Stream RAG (parallel tool call) are key instances. This pattern is critical in real-time interaction scenarios.

\item \textbf{Data-driven Orchestration Pattern}: The agent learns its planning and tool-use policy from data, rather than relying on static prompts or rules. AuTAgent is the archetype, using RL to learn when and which tools to use, overcoming the brittleness of hand-crafted heuristics.

\item \textbf{Asynchronous Multi-Agent Pattern}: Separate agents run in parallel, decoupling latency-critical components from heavy tools. VoxMind exemplifies this with an auxiliary agent that executes retrieval asynchronously while the main model generates a response.
\end{itemize}

Recent trends highlight three key developments: proactive streaming patterns (SHANKS, LTS-VoiceAgent, Stream RAG) are breaking the ``listen-then-act'' assumption; RL-based learned orchestration (AuTAgent) is surpassing prompt-based tool selection; and asynchronous multi-agent designs (VoxMind) decouple tool execution from response generation, enabling hybrid efficiencies. These patterns provide a toolkit for designing future agentic audio reasoning systems.

\subsection{Discussion}

Table~\ref{tab:tradeoffs} summarizes the core trade-offs between the two dominant paradigms. Predefined workflows remain the choice for high-reliability applications where interpretability and traceability are critical. Their streaming variants (SHANKS, LTS) now achieve low latency while retaining deterministic behavior. Dynamic tool-calling agents offer superior flexibility for open-ended tasks, and the shift from prompt-based orchestration (AudioToolAgent) to learned policies (AuTAgent) signals a maturing of the paradigm.


\section{Evaluations}
\label{sec: Evaluations}

Evaluating audio reasoning models requires assessing not only final answer correctness but also whether the underlying reasoning is genuinely grounded in acoustic evidence. In this section, we survey the benchmarks, metrics, and protocols used to evaluate models across the Audio-to-Text and Audio-to-Speech reasoning paradigms.

\subsection{Audio-to-Text Evaluation}

Audio-to-text evaluation aims to assess whether models can correctly understand spoken inputs and produce textual outputs that reflect both the semantic content and the underlying acoustic cues. Compared with conventional speech recognition, the focus here is not limited to recovering lexical content, but extends to reasoning over speech in a way that is robust to acoustic variability and sensitive to non-textual information when necessary. From this perspective, audio-to-text evaluation can be organized around two aspects: the tested reasoning capability and the metrics.

\subsubsection{Reasoning Capability}

A central goal of audio-to-text evaluation is to assess whether a model can reason over spoken input rather than merely transcribe it. Existing settings can be divided into content-based reasoning and acoustic-based reasoning, depending on whether inference relies on linguistic content or acoustic signals.

Content-based reasoning includes tasks where evidence comes from semantic content, such as spoken QA~\cite{chen2024voicebench}, dialogue understanding~\cite{gosai2025audio,zhang2025mtr}, and instruction following~\cite{lu2025speech, wang2025inserter}. The model needs to recover meaning and reason over it to produce correct responses. Instruction following in speech is also a form of content-based reasoning, as it requires understanding and executing semantic intent. Recent settings~\cite{lu2025speech, wang2025inserter,chen2024voicebench} further introduce content-preserving acoustic variations, such as emotion, speaking rate, or background noise. These aim not to test reasoning over acoustic cues, but to evaluate whether content-based reasoning remains stable under such perturbations.

In contrast, acoustic-based reasoning~\cite{sakshi2024mmau, mmsu, ma2025mmar} involves tasks where correct inference depends on information not reducible to text, including prosody, emotion, speaker traits, emphasis, hesitation, and other paralinguistic signals. Text transcripts alone are insufficient, and the model must use acoustic evidence. Thus, acoustic-based reasoning more directly tests whether models leverage speech-specific information, rather than relying on implicit text-only processing. This distinction is key to understanding whether systems truly reason over speech or mainly over textual abstractions.

\subsubsection{Evaluation Metrics}

Evaluation of audio-to-text tasks typically includes both objective metrics, which measure correctness against references, and subjective evaluation, which assesses quality in open-ended settings.

\textbf{ACC-Metrics.} A common approach is to use accuracy-based metrics from language model benchmarks~\cite{sakshi2024mmau, mmsu, ma2025mmar,mmau}, especially when outputs are discrete or can be normalized into a small set. This is typical in multiple-choice QA, classification, and closed-ended reasoning. Common metrics include accuracy, F1, and, in retrieval or imbalanced settings, mean average precision (mAP). These directly measure agreement with ground truth and suit controlled reasoning evaluation. Many spoken reasoning benchmarks adopt this paradigm for clear comparison and reduced ambiguity.

\textbf{Text-overlap Metrics.} For free-form outputs, exact match can be too strict, as correct answers may vary in wording. Text-overlap metrics such as BLEU, ROUGE, and METEOR measure similarity between outputs and references. They are useful for structured generation tasks like summarization, short QA, or captioning, where references exist and surface similarity is informative. However, they poorly capture reasoning quality, as responses may be semantically correct but lexically different, or similar yet incorrect.

\textbf{Human Evaluation.} Subjective evaluation is important for open-ended tasks where multiple responses are valid or correctness depends on nuanced interpretation. Evaluators assess correctness, coherence, relevance, faithfulness to the audio, and reasoning quality. It is especially valuable for long-form responses, dialogue, and explanations, where objective metrics fall short. Despite high cost and limited scalability, it is still considered a highly reliable evaluation approach.

\textbf{LLM-based Evaluation.} To improve scalability, recent work uses LLMs as evaluators~\cite{yang2024air,chen2024voicebench,li2026wavbench}. The evaluator model scores or compares outputs based on criteria such as correctness, completeness, or consistency with audio-derived context. This approach is attractive for open-ended reasoning due to flexibility and lower cost than human annotation. However, reliability depends on prompt design, bias, and the model’s ability to assess audio-grounded reasoning, so it is best viewed as a scalable proxy rather than a replacement for human evaluation.

\subsection{Audio-to-Speech Evaluation}

Audio-to-speech evaluation assesses whether models can generate spoken responses that are acoustically natural, semantically correct, and contextually appropriate. While all audio-to-text evaluation metrics apply equally here, audio-to-speech evaluation goes further to encompass interaction ability, response alignment, and real-time behavior that are unique to spoken dialogue systems.

\textbf{Dialogue-oriented ability.} A key aspect is supporting coherent spoken interaction, including maintaining consistency across turns, adapting to context, and generating appropriate replies. It also requires sensitivity to affective and pragmatic cues such as emotion, emphasis, and speaking style. In advanced settings, evaluation includes full-duplex interaction, handling overlap, turn-taking, and interruptions.

\textbf{Speech quality.} This evaluates whether generated audio is natural and intelligible, considering pronunciation, prosody, and absence of artifacts like distortion or unnatural pauses. These remain essential in AR-LLMs, where speech depends on dynamic content and reasoning.

\textbf{Response alignment.} This measures whether speech reflects semantic content and contextual intent, including correctness, consistency with dialogue history, and alignment between meaning and paralinguistic expression. For example, emotion and prosody should match the intended message. Misalignment may produce fluent but inappropriate responses.

\textbf{Real-time interaction.} This evaluates performance under latency and streaming constraints, including response delay, incremental generation, and stability. Low latency supports natural flow, while robust streaming is needed for long-form spoken dialogue.

\textbf{Subjective evaluation.} Perceptual quality and interaction effectiveness are inherently difficult to quantify with automatic metrics, making subjective assessment indispensable. Mean Opinion Score (MOS) is the standard measure for speech naturalness and quality. Other human or model-based evaluation captures higher-level attributes such as coherence, emotional appropriateness, and conversational effectiveness.

\begin{table*}[htbp]
\caption{Overview of Reasoning Benchmarks for SLMs. Open and closed denote open-ended and closed-ended evaluation types, respectively.}
\label{tab:benchmarks}
\begin{center}
\small
\setlength{\tabcolsep}{4pt}

\begin{tabularx}{\textwidth}{l l X l l l l}
\toprule
\textbf{Benchmark} & \textbf{Domain} & \textbf{Audio Type} & \textbf{Eval Type} & \textbf{Size} & \textbf{\#Tasks} & \textbf{Year} \\
\midrule
MMAU~\cite{mmau}
& General Audio Reasoning
& Speech, Sound, Music
& Closed
& 10K
& 27
& 2024 \\

MMAU-Pro~\cite{kumar2025mmau}
& General Audio Reasoning
& Speech, Sound, Music
& Closed
& 0.5K 
& 49
& 2025 \\

MMAR~\cite{ma2025mmar}
& General Audio Reasoning
& Speech, Sound, Music
& Closed
& 1K
& 16
& 2025 \\

MMSU~\cite{mmsu}
& Speech-focused Reasoning
& Speech
& Closed
& 5K
& 47
& 2025 \\

VoxEval~\cite{voxeval}
& Speech-focused Reasoning
& Speech
& Closed
& 14K
& 56
& 2025 \\


CMDAR~\cite{li2025mdar}
& Chinese Audio Reasoning
& Speech, Sound
& Closed, Open
& 3K
& 5
& 2025 \\

Spoken-MQA~\cite{wei2025towards}
& Spoken Mathematical Reasoning
& Speech
& Closed, Open
& 3K
& 5
& 2025 \\

WildSpeech-Bench~\cite{zhang2025wildspeech}
& End-to-end Interaction
& Speech
& Open
& 1.1K
& 5
& 2025 \\

URO-Bench~\cite{yan2025uro}
& End-to-end Interaction
& Speech
& Closed, Open
& 5K
& 20
& 2025 \\

VoiceAgentBench~\cite{jain2025voiceagentbench}
& Multi-turn Tool Use \& Reasoning
& Speech
& Closed
& 6K
& 6
& 2025 \\

SAKURA~\cite{yang2025sakura}
& Multi-hop audio reasoning
& Speech, Sound
& Closed
& 4K
& 4
& 2025 \\


WavBench~\cite{li2026wavbench}
& End-to-end Interaction
& Speech
& Closed, Open
& 2K
& 5 (scenarios)
& 2026 \\

$\tau$-Voice~\cite{ray2026tau}
& End-to-end Interaction
& Speech
& Closed
& 0.3K
& 3 (domains)
& 2026 \\

AudioRAG~\cite{lin2026audiorag}
& Audio Retrieval-based Reasoning  
& Speech, Sound, Music
& Closed
& 8K
& 3 (scenarios)
& 2026 \\

Full-Duplex-Bench-v3~\cite{fdb_v3_0}
& Multi-turn Tool Use \& Reasoning
& Speech
& Closed
& 0.1k
& 4
& 2026 \\

\bottomrule
\end{tabularx}
\end{center}
\end{table*}

\section{Challenges and Future Directions}
\label{sec: Challenges and Future Directions}

\subsection{Reliability of Synthesized Reasoning Data}
The development of robust audio reasoning models is fundamentally constrained by the scarcity of high-quality, domain-specific training data. To circumvent the prohibitive costs of human annotation, a common paradigm involves leveraging powerful text-only LLMs as data engines to synthesize reasoning trajectories based on acoustic metadata, such as transcripts or sound event tags. However, relying exclusively on text-based LLMs to construct these reasoning chains introduces significant reliability concerns. Because this synthetic data is derived from discrete textual approximations rather than the continuous acoustic signal, there is no guarantee that the resulting reasoning chains are genuinely grounded in the actual audio input. Consequently, acquiring reasoning trajectories that faithfully capture complex, continuous acoustic nuances—without incurring immense human annotation costs—remains a critical bottleneck for the field.

\subsection{Modality Hallucination and Shortcut Learning}
A fundamental architectural challenge in current LALMs stems from an inherent modality mismatch during initialization. As most LALMs inherit their reasoning capabilities from backbone models trained predominantly on text, they frequently exhibit shortcut learning or modality hallucination. Specifically, models often default to text-surrogate reasoning \cite{step-audio-r1}, where they rely on discrete audio captions or intermediate text representations to perform logical deduction rather than grounding their analysis in the continuous acoustic signal. Consequently, rather than executing genuine acoustic reasoning, these models sometimes bypass the auditory input entirely, relying excessively on the strong semantic priors of their language backbone to infer answers. Mitigating this over-reliance and ensuring that models explicitly anchor their reasoning in raw auditory cues is essential for developing faithful audio reasoning systems.

\subsection{The Accuracy vs. Latency Trade-Off}
For the Audio-to-Speech task, researchers face an intrinsic trade-off between reasoning accuracy and the strict latency constraints required for real-time human-computer interaction. Generating extensive, multi-step reasoning chains inevitably delays the system's acoustic response, disrupting the natural flow of conversation. To mitigate this, emerging architectural paradigms---such as ``thinking while listening'' or ``thinking while speaking''---attempt to parallelize the reasoning computation with the user's input or the system's output. However, an observable performance gap persists between these streaming approaches and their offline counterparts \cite{chiang2026stitch,chiang2025shanks}. Models performing latency-constrained reasoning currently struggle to match the high accuracy and rigorous deduction achieved by unconstrained audio reasoning or pure text-based reasoning systems. Overcoming this performance degradation while maintaining the ultra-low latency necessary for seamless, full-duplex communication remains a critical challenge.

\subsection{Long-Context Audio Reasoning}
Current methodologies in audio reasoning have primarily demonstrated efficacy on relatively short, well-defined audio clips. However, scaling these reasoning capabilities to process continuous, long-form audio streams—such as meetings, podcasts, or environmental recordings lasting several minutes—introduces significant computational and cognitive uncertainties. Feeding uncompressed, temporally dense acoustic features directly into a model often results in computationally prohibitive sequence lengths. Furthermore, it remains empirically unclear whether current models can successfully maintain precise audio perception and context retention over these extended durations. The ability to execute multi-hop reasoning over lengthy acoustic scenes, where a model must recall and integrate distinct audio events separated by minutes of context, is an open question that requires dedicated benchmarking and architectural innovation.

\subsection{Shifting Toward Foundational Pre-Training for Reasoning}
The prevailing paradigm for eliciting audio reasoning capabilities heavily relies on post-training interventions. Contemporary approaches typically apply SFT or RL to models that have already undergone basic modality alignment. While effective to a degree, this post-training focus raises questions about the foundational limits of the underlying model capabilities. A promising yet underexplored direction is the integration of reasoning-oriented objectives during the pre-training or mid-training phases. By cultivating deep, intrinsic alignment and logical structuring earlier in the training pipeline---potentially through large-scale, joint audio-text predictive objectives---researchers might establish a more robust cognitive foundation. It remains to be investigated whether establishing this structural groundwork during pre-training can make subsequent post-training reasoning enhancements significantly more effective and native to the audio modality.

\section{Conclusion}

In this survey, we review the emerging field of audio reasoning, highlighting key challenges such as modality mismatch, data scarcity, shortcut learning, and latency constraints. We organize existing approaches into Audio-to-Text and Audio-to-Speech, with an extension to the Audio-Visual domain. We examine techniques, including Chain-of-Thought reasoning and agentic frameworks, and how these frameworks meet the interaction latency constraint. Through our analysis, we identified several critical problems in current works, notably their tendency for modality hallucination, where models bypass auditory cues to rely on text-surrogate reasoning, the reliability concerns of using text-only LLMs to synthesize acoustic reasoning data, an unresolved trade-off between deep reasoning accuracy and real-time interaction latency, and uncertainties in processing long-context continuous audio streams. Overall, we provide a unified perspective and outline future directions for building robust, real-time, and genuinely audio-grounded reasoning systems.



%
\bibliographystyle{IEEEtran}
\bibliography{reference_short}

\begin{IEEEbiography}
 [{\includegraphics[width=1in,height=1.25in,clip,keepaspectratio]{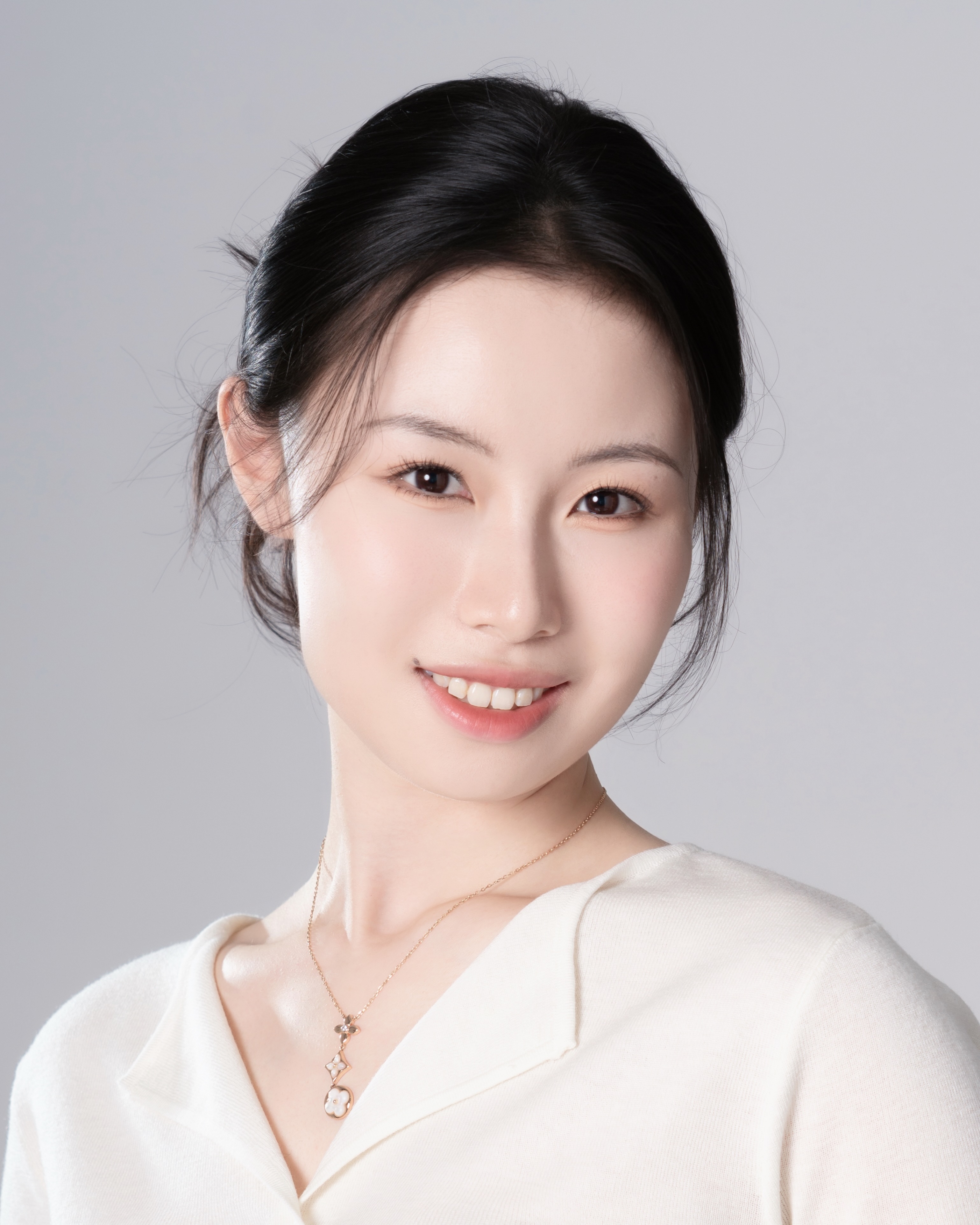}}]
 {Zhihan Guo} received the B.Eng. degree from Beijing Institute of Technology. She is currently a Ph.D. student at the Department of Computer Science and Engineering of the Chinese University of Hong Kong (CUHK), under the supervision of Prof. Irwin King. Her research interests include long text generation and agentic AI.
\end{IEEEbiography}

\vspace{-1.5em}

\begin{IEEEbiography}
 [{\includegraphics[width=1in,height=1.25in,clip,keepaspectratio]{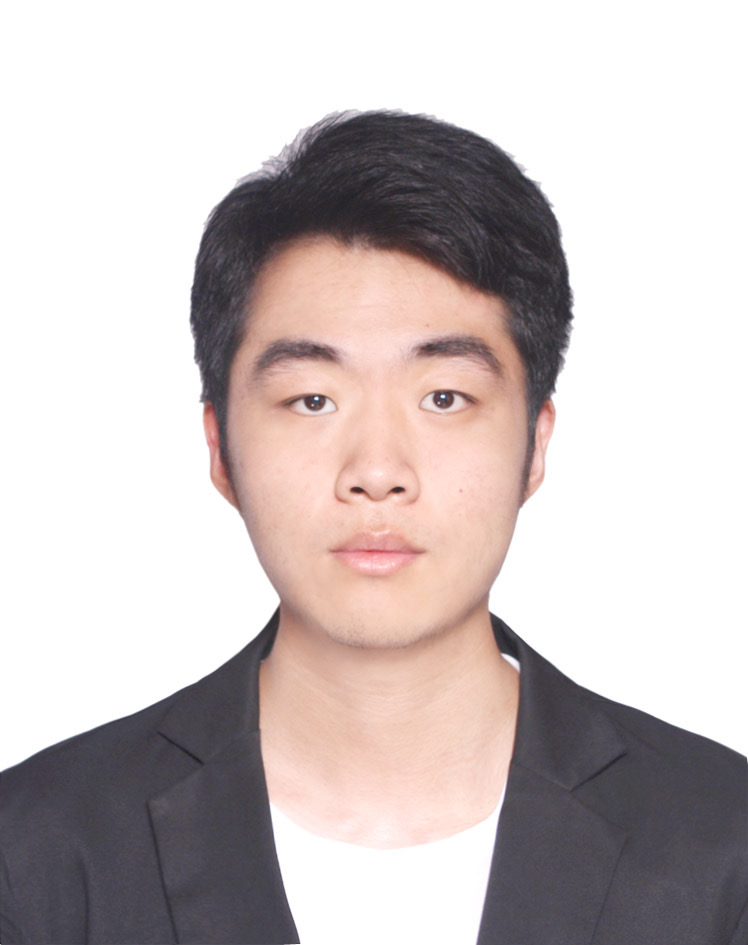}}]
 {Wenqian Cui} received the B.Eng. degree from Beijing University of Posts and Telecommunications. He is currently a Ph.D. student at the Department of Computer Science and Engineering of the Chinese University of Hong Kong (CUHK), under the supervision of Prof. Irwin King. His research interests include multimodal large language models and AI for audio.
\end{IEEEbiography}

\vspace{-1.5em}

\begin{IEEEbiography}
 [{\includegraphics[width=1in,height=1.25in,clip,keepaspectratio]{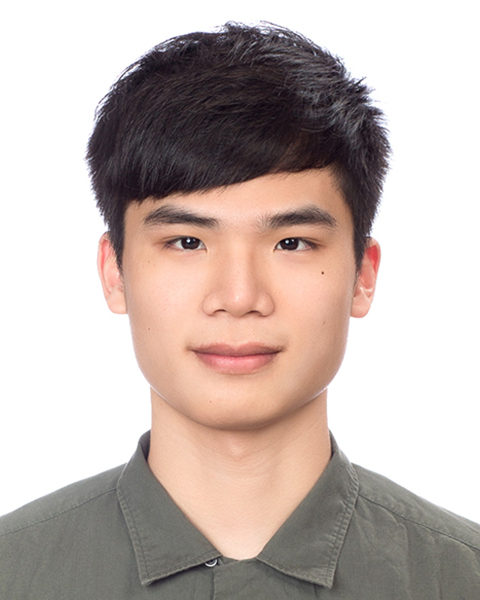}}]
 {Guan-Ting Lin} received the Ph.D. degree in Communication Engineering from National Taiwan University, under the supervision of Prof. Hung-yi Lee. He serves as a reviewer for ICLR, NeurIPS, ACL, EMNLP, NAACL, ICASSP and etc. His research interests include spoken language models, spoken dialogue system, and full-duplex interaction.
\end{IEEEbiography}

\vspace{-1.5em}

\begin{IEEEbiography}
 [{\includegraphics[width=1in,height=1.25in,clip,keepaspectratio]{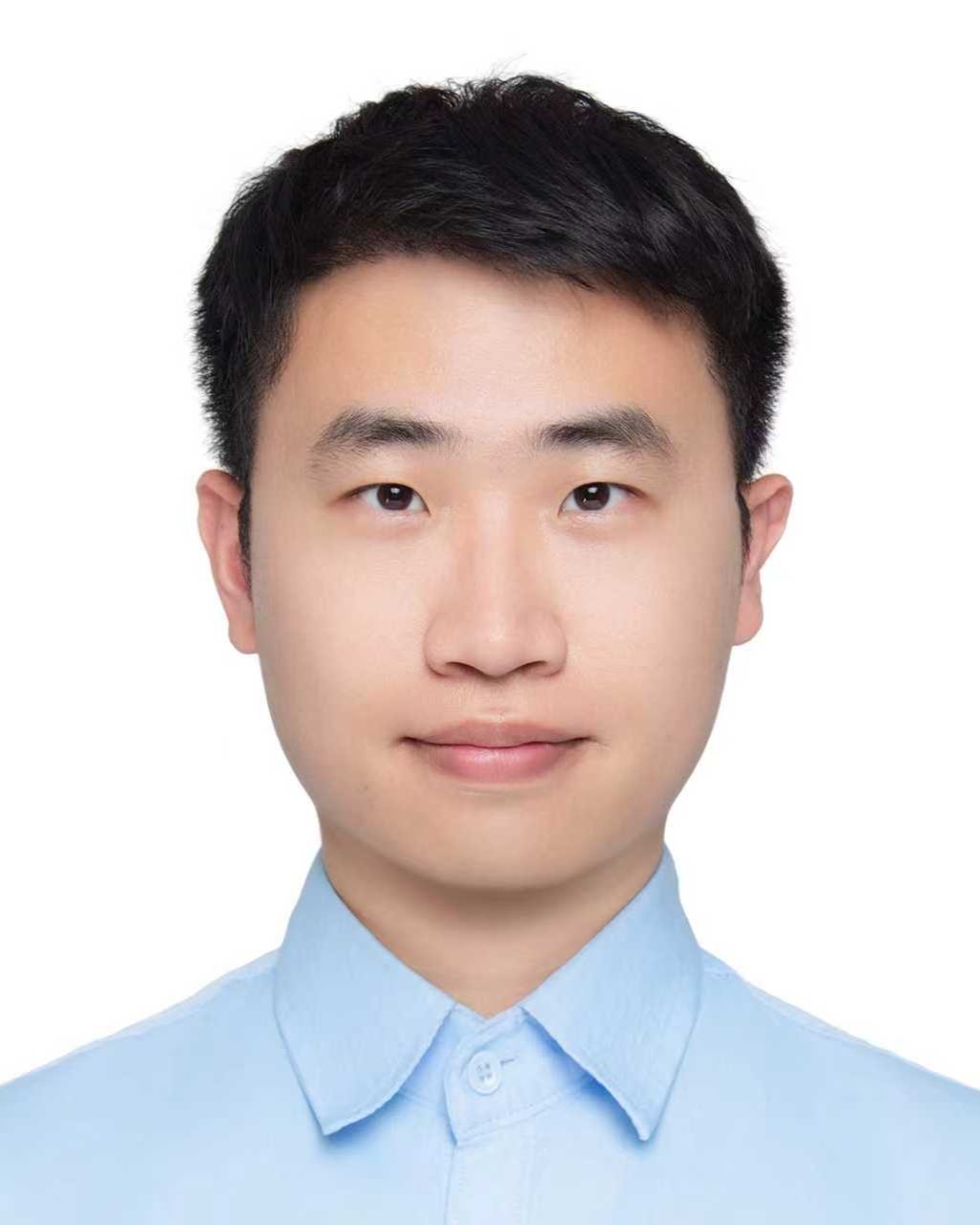}}]
{Daxin Tan} received the Ph.D. degree from the Department of Electronic Engineering of The Chinese University of Hong Kong, supervised by Prof. Tan Lee, and the B.E. degree from Tsinghua University. His research interests include speech representation learning, text-to-speech synthesis, and multi-modal large language models.
\end{IEEEbiography}

\vspace{-1.5em}

\begin{IEEEbiography}
 [{\includegraphics[width=1in,height=1.25in,clip,keepaspectratio]{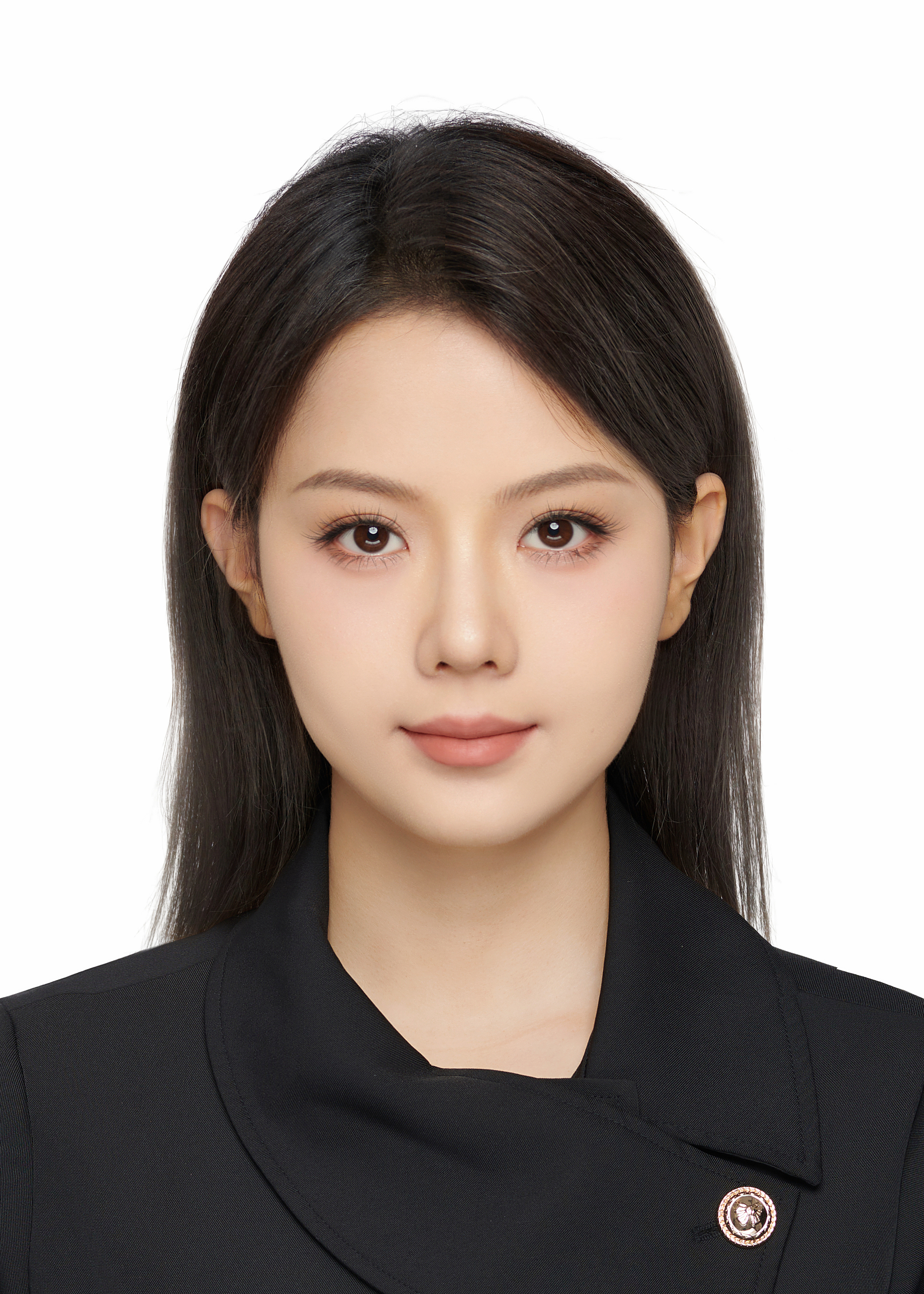}}]
 {Jingyao Li} received the B.Eng. degree from Xi'an Jiaotong University. She is currently a Ph.D. student at Department of Computer Science and Engineering of the Chinese University of Hong Kong (CUHK), under the supervision of Prof. Jiaya Jia. Her research interests include training of foundation large language models.
\end{IEEEbiography}

\vspace{-1.5em}

\begin{IEEEbiography}
 [{\includegraphics[width=1in,height=1.25in,clip,keepaspectratio]{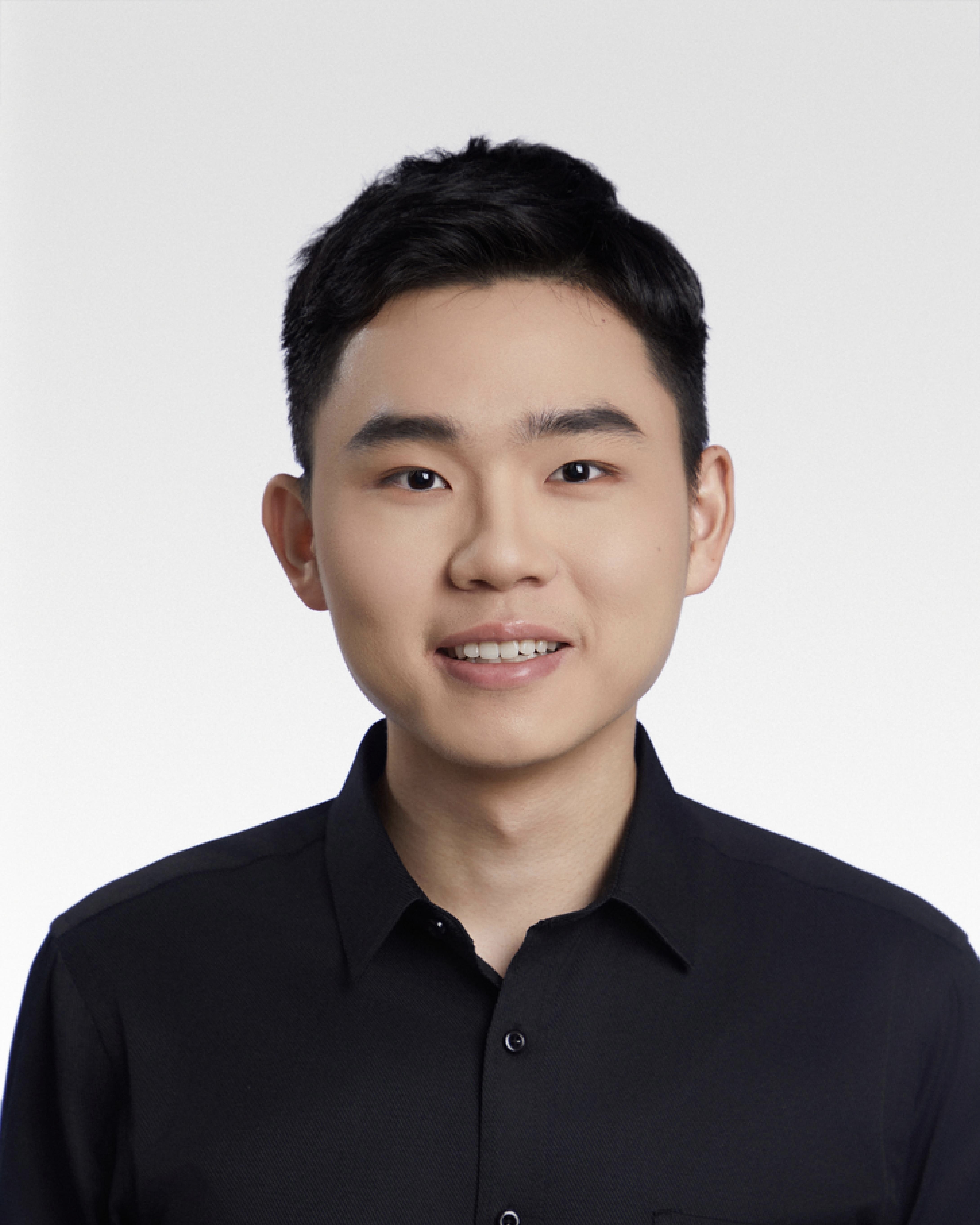}}]
 {Qiyong Zheng} received the B.Sc. degree from Shenzhen University. He is currently an MPhil student at the Department of Computer Science and Engineering of the Chinese University of Hong Kong (CUHK), under the supervision of Prof. Irwin King. His research interest lies in post-training for large language models.
\end{IEEEbiography}

\vspace{-1.5em}

\begin{IEEEbiography}
 [{\includegraphics[width=1in,height=1.25in,clip,keepaspectratio]{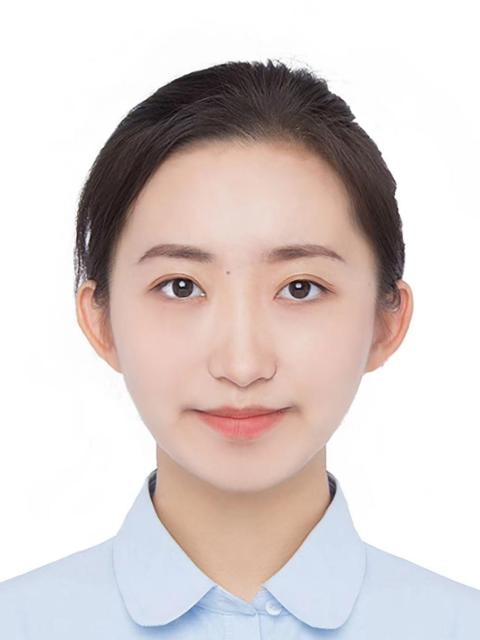}}]
{Dingdong Wang}received the B.Sc. degree from The Chinese University of Hong Kong (CUHK). She is currently a Ph.D. student in the Department of Systems Engineering and Engineering Management at CUHK, under the supervision of Prof. Helen Meng. Her research interests include multimodal large language models and AI for audio.
\end{IEEEbiography}

\vspace{-1.5em}

\begin{IEEEbiography}
 [{\includegraphics[width=1in,height=1.25in,clip,keepaspectratio]{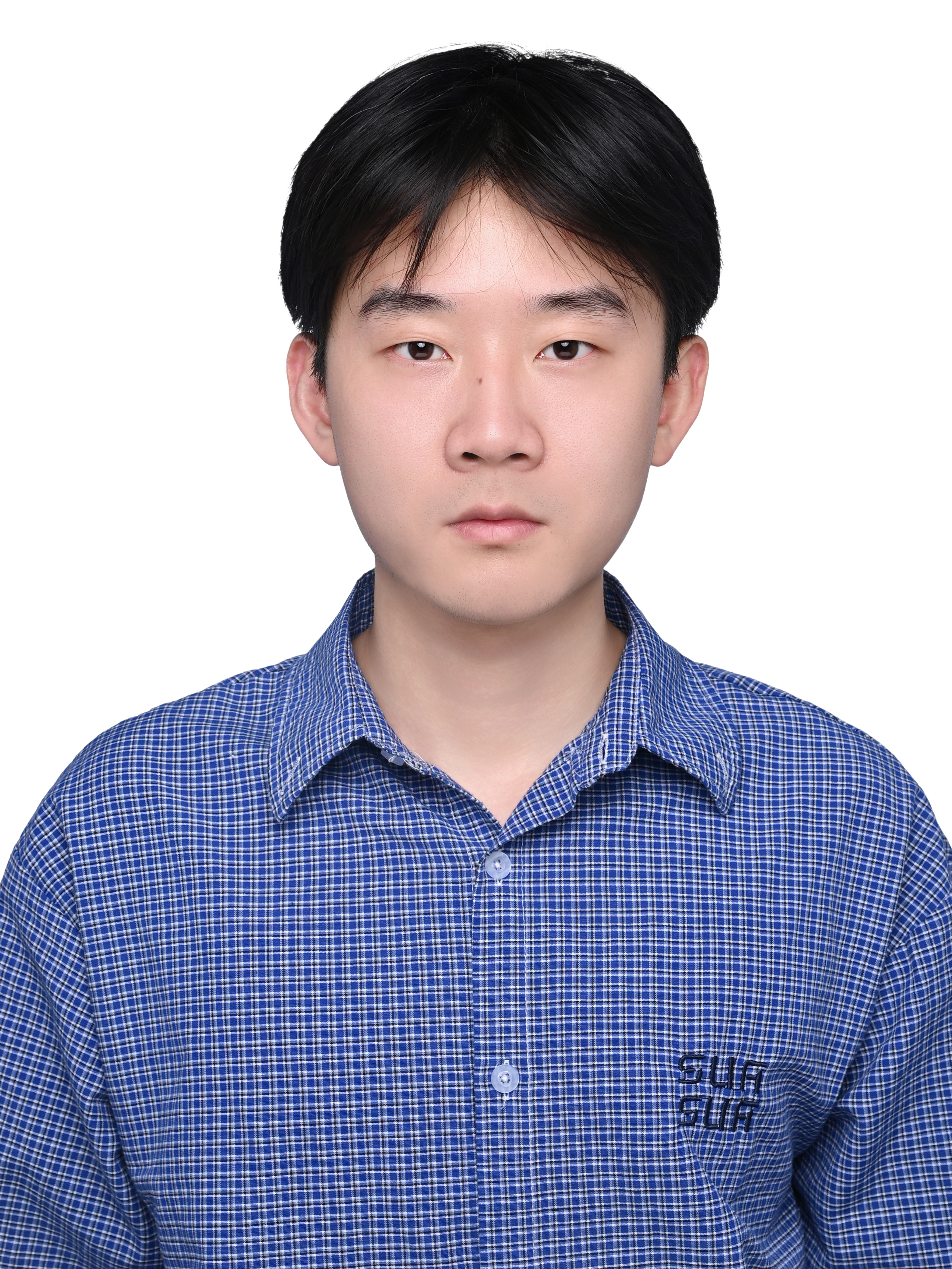}}]
{Jing Xiong} is currently a PhD student at the University of Hong Kong, supervised by Prof. Yi Huang and Prof. Lingpeng Kong. He has published multiple papers in top-tier conferences and journals, including ICLR, ICML, NeurIPS, ACL, EMNLP, and TMLR. His research primarily focuses on efficient language model inference and automated theorem proving.
\end{IEEEbiography}

\vspace{-1.5em}

\begin{IEEEbiography}
 [{\includegraphics[width=1in,height=1.25in,clip,keepaspectratio]{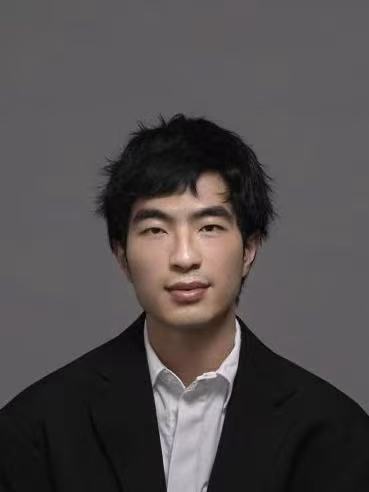}}]
 {Han Shi} is a research scientist at Huawei Foundation Model Department. He obtained his Ph.D. degree from Department of Computer Science and Engineering (CSE) in HKUST, supervised by Prof. James T. Kwok. Before his Ph.D. journey, he obtained his Bachelor degree in Electronic Engineering from Tsinghua University. His current research interests are Generative Models, Neural Architecture Design and Automated Machine Learning (AutoML).
\end{IEEEbiography}

\vspace{-1.5em}

\begin{IEEEbiography}
[{\includegraphics[width=1in,height=1.25in,clip,keepaspectratio]{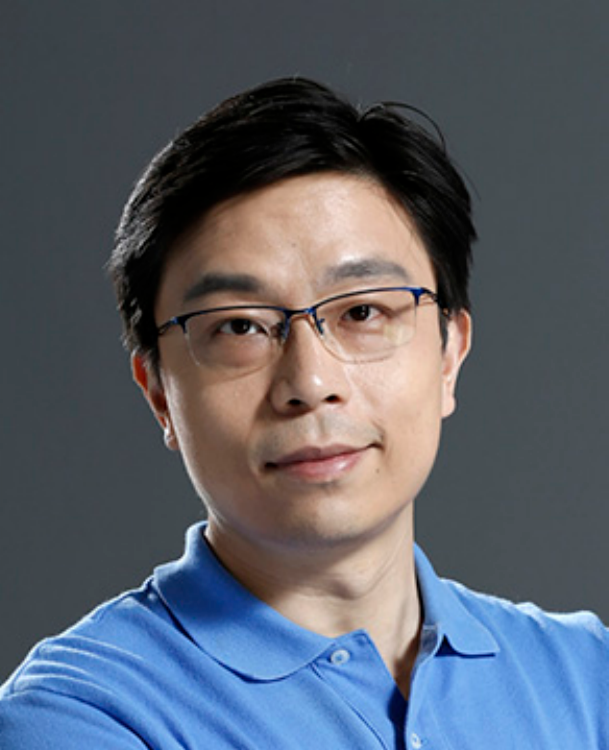}}]{Jiaya Jia} received the Ph.D.~degree in Computer Science from Hong Kong University of Science and Technology in 2004 and is currently a chair professor in Department of Computer Science and Engineering at the Hong Kong University of Science and Technology (HKUST). He assumes the position of Associate Editor-in-Chief of IEEE Transactions on Pattern Analysis and Machine Intelligence (TPAMI) and is in the editorial board of International Journal of Computer Vision (IJCV). He continuously served as area chairs for ICCV, CVPR, AAAI, ECCV, and several other conferences for the organization. He was on program committees of major conferences in graphics and computational imaging, including ICCP, SIGGRAPH, and SIGGRAPH Asia. He is a Fellow of the IEEE. 
\end{IEEEbiography}

\vspace{-1.5em}

\begin{IEEEbiography}
 [{\includegraphics[width=1in,height=1.25in,clip,keepaspectratio]{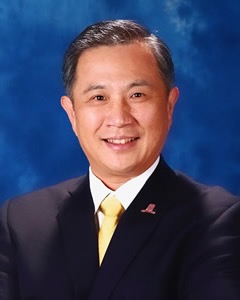}}]
 {Irwin King} is a globally recognized scholar in the field of machine intelligence, currently serving as the Pro-Vice-Chancellor (Education) and Professor at the Department of Computer Science \& Engineering at The Chinese University of Hong Kong. His extensive research interests encompass a wide range of areas, including trustworthy AI, machine learning, social computing, AI, and data mining. Professor King is a Fellow of esteemed societies and associations, such as the ACM, IEEE, AAAI, and INNS. 
\end{IEEEbiography}
\newpage

 




\vfill

\end{document}